\documentclass[a4paper,11t]{article}
\usepackage[top=2.54cm,bottom=2.54cm, left=2cm, right=2cm]{geometry}

\usepackage{amsmath}
\usepackage{amsfonts}
\usepackage{amssymb}
\usepackage{graphicx}
\usepackage{bm}
\usepackage{enumerate}
\usepackage{color}
\usepackage[sort,comma]{natbib}
\usepackage{float}
\usepackage[colorlinks=true, allcolors=blue]{hyperref}
\usepackage{changepage}
\usepackage{dirtytalk}
\usepackage{pdflscape}

\title{Circular and Spherical Projected Cauchy Distributions: A Novel Framework for Circular and Directional Data Modeling}
\author{Michail Tsagris and Omar Alzeley\\
\\
Department of Economics, University of Crete, \\
Gallos Campus, Rethymnon, Greece \\ 
\href{mailto:mtsagris@uoc.gr}{mtsagris@uoc.gr} \\
Department of Mathematics, Umm Al-Qura University, \\
Al-Qunfudah University College, Saudi Arabia \\ 
\href{mailto:oazeley@uqu.edu.sa}{oazeley@uqu.edu.sa} \\
}
\begin{document}

\maketitle

\begin{center}
\textbf{Abstract}
\end{center}
We introduce a novel family of projected distributions on the circle and the sphere, namely the circular and spherical projected Cauchy distributions, as promising alternatives for modelling circular and spherical data. The circular distribution encompasses the wrapped Cauchy distribution as a special case, while featuring a more convenient parameterisation. We also propose a generalised wrapped Cauchy distribution that includes an extra parameter, enhancing the fit of the distribution. In the spherical context, we impose two conditions on the scatter matrix of the Cauchy distribution, resulting in an elliptically symmetric distribution. Our projected distributions exhibit attractive properties, such as a closed-form normalising constant and straightforward random value generation. The distribution parameters can be estimated using maximum likelihood, and we assess their bias through numerical studies. Further, we compare our proposed distributions to existing models with real datasets, demonstrating equal or superior fitting both with and without covariates.  \\
\\
\textbf{Keywords:} Directional data; Cauchy distribution; elliptically symmetric distribution
\\
\\
MSC: 62H11, 62H10
\section{Introduction}
Directional data refers to multivariate data with a unit norm, whose sample space can be expressed as:
\begin{eqnarray*}
\mathbb{S}^{d-1}=\left\lbrace {\bf x} \in \mathbb{R}^d \bigg\vert \left|\left|{\bf x}\right|\right|=1 \right\rbrace,
\end{eqnarray*}

where $\left|\left|.\right|\right|$ denotes the Euclidean norm. Circular data, when $d=2$, lie on a circle, whereas spherical data, when $d=3$, lie on a sphere. Circular data are encountered in various disciplines, such as political sciences \citep{gill2010}, criminology \citep{shirota2017}, biology \citep{landler2018}, ecology \citep{horne2007}, and astronomy \citep{soler2019}, to name but a few. Spherical data, on the other hand, are encountered in fields such as geology \citep{chang1986}, environmental sciences \citep{heaton2014}, image analysis \citep{straub2015}, robotics \citep{bullock2014}, and space \citep{kent2016}.

A vast array of circular distributions have been reported in the literature, with the earliest being the von Mises distribution \citep{von1918}, investigated by \cite[pg.~36--41]{mardia2000}. This distribution emerges as the conditional distribution of a bivariate normal random vector with a specific mean vector and identity covariance matrix, given that the vector lies on the unit circle. Over the years, various generalisations of this distribution have been proposed \citep{gatto2007,kim2013,dietrich2017}, with additional circular distributions introduced by \cite{pewsey2000}, \cite{jones2005}, \cite{abe2011}, and \cite{jones2012}. Wrapped distributions represent another class of distributions arising from wrapping a univariate random vector on the circle. This category includes the wrapped $t$ family of distributions \citep{pewsey2007}, the wrapped stable family \citep{pewsey2008}, the wrapped normal, and wrapped Cauchy (WC) distributions \citep[pg.~50--52]{mardia2000}, as well as their extensions \citep{kato2010,kato2013}. Finally, a less explored type of distributions, known as projected distributions, originate from the distribution of a multivariate random vector projected onto a circle. The bivariate projected normal (PN) distribution (\citealp[pg.~52]{mardia1972}, \citealp{presnell1998}) is likely the only distribution of this kind.

Incorporating more than two parameters in circular probability distributions is a common practice intended to enhance flexibility and better capture skewed data. The PN distribution, an exception to this issue, encompasses multiple parameters and has been employed to model various data types, including skewed data. To estimate the scatter matrix, in the general PN distribution, \cite{nunez2005,wang2013,hernandez2017} adopted the Bayesian stance.

Numerous spherical and hyper-spherical distributions have been proposed over time, with the von Mises-Fisher \citep{fisher1953} and PN \citep{kendall1974} distributions being among the earliest and most prevalent, while the spherical Cauchy (SC) \citep{kato2020} is a more recent proposition. However, these distributions assume rotational symmetry, which may restrict their applicability in certain scenarios. To mitigate this constraint, \cite{kent1982} introduced an elliptically symmetric distribution that relaxes the rotational symmetry assumption. This distribution constitutes a special case of the Fisher-Bingham distribution \citep{mardia1975}, and has proven valuable for modelling more sophisticated data structures. More recently, \cite{paine2018} proposed the elliptically symmetric angular Gaussian (ESAG) distribution, which emerges by projecting the Gaussian distribution onto the sphere and imposing two conditions on the covariance matrix.

In this study, we introduce the circular and spherical projected Cauchy distributions, representing noteworthy additions to the family of directional distributions as they provide an alternative perspective on the Cauchy distribution. Notable characteristics of these new distributions include their closed-form normalising constant and a simplicity of simulation, rendering them particularly attractive for modelling directional data. By projecting the bivariate Cauchy distribution onto the circle, we derive a circular distribution. When the scatter matrix is equivalent to the identity matrix, the resulting distribution corresponds to the WC distribution, albeit with a more convenient parameterisation. This significant outcome enables us to associate the circular projected Cauchy with a well-established distribution. Subsequently, in line with \cite{paine2018}, we impose a location-constrained scatter matrix, offering enhanced flexibility and facilitating more realistic modelling of circular data. The spherical projected Cauchy distribution is also of interest because after imposing the same conditions as described in \cite{paine2018}, it yields an elliptically symmetric distribution for spherical data, a property rarely encountered in the literature. This characteristic is of paramount importance with regard to spherical data given the limited number of distributions with this property documented in the literature.

We present the proposed circular and spherical projected Cauchy distributions, including the regression setting, in Sections \ref{circle} and \ref{sphere}, respectively. To compare the performance of the various forms of the proposed projected Cauchy distributions with one another, and indeed with alternative distributions, we conduct simulation studies in Section \ref{sims}. We demonstrate the performance of the proposed distributions and selective competing distributions using examples of real data in Section \ref{real}. Section \ref{conc} offers concluding remarks on the paper.

\section{The Circular Projected Cauchy Distribution and its Special Cases} \label{circle}
Suppose a $d$-dimensional random variable ${\bf X}$ follows some multivariate distribution defined over $\mathbb{R}^d$ and we project it onto the circle/sphere/hyper-sphere, ${\bf Y}=\frac{{\bf X}}{r}$, where $r=\left|\left|{\bf X}\right|\right|$. The marginal distribution of ${\bf Y}$, which is of interest, is obtained by integrating out $r$ over the positive line
\begin{eqnarray} \label{projden}
f({\bf y})=\int_0^{\infty}r^{d-1}f(r{\bf y})dr.
\end{eqnarray}

The probability density function of the bivariate Cauchy distribution, with some location vector $\pmb{\mu}$ and scatter matrix $\pmb{\Sigma}$, is given by
\begin{eqnarray} \label{bivt2}
f({\bf x})=\frac{1}{2\pi|\pmb{\Sigma}|^{1/2}}\left[1 + \left({\bf x}-\pmb{\mu}\right)^\top\pmb{\Sigma}^{-1}\left({\bf x}-\pmb{\mu}\right)\right]^{-3/2}.
\end{eqnarray}

By substituting (\ref{bivt2}) into (\ref{projden}) and evaluating the integral, we arrive at a new distribution on the circle, termed the Circular Projected Cauchy (CPC) distribution.
\begin{eqnarray} \label{pc}
f({\bf y}) &=& \int_0^{\infty}\frac{r}{2\pi|\pmb{\Sigma}|^{1/2}}\left(1 + r^2{\bf y}^\top\pmb{\Sigma}^{-1}{\bf y}-2r{\bf y}^\top\pmb{\Sigma}^{-1}\pmb{\mu}+\pmb{\mu}^\top\pmb{\Sigma}^{-1}\pmb{\mu}\right)^{-3/2}dr \nonumber \\
&=& \frac{1}{2\pi|\pmb{\Sigma}|^{1/2}\left(B\sqrt{\Gamma^2+1}-A\sqrt{B}\right)},
\end{eqnarray}
where 
\begin{subequations}
\begin{align}
A={\bf y}^\top\pmb{\Sigma}^{-1}\pmb{\mu} \label{A} \\
B={\bf y}^\top\pmb{\Sigma}^{-1}{\bf y} \label{B} \\
\Gamma^2=\pmb{\mu}^\top\pmb{\Sigma}^{-1}\pmb{\mu} \label{G2}
\end{align}
\end{subequations}
It is important to note that ${\bf y} \in \mathbb{S}^1$, while $\bm{\mu} \in \mathbb{R}^2$.

\subsection{The WC as a Special Case of the CPC Distribution}
The difficulty with the CPC is the excessive number of associated parameters and thus we assume $\pmb{\Sigma}={\bf I}_2$, resulting in the following representation:
\begin{eqnarray} \label{cipc}
f({\bf y}) =\dfrac{1}{2\pi\left(\sqrt{\gamma^2+1}-\alpha\right)},
\end{eqnarray}
where 
\begin{subequations}
\begin{align}
\alpha={\bf y}^\top\pmb{\mu} \label{a} \\
\gamma=\|\pmb{\mu}\|. \label{g} 
\end{align}
\end{subequations}
The density in (\ref{cipc}) can also be written as
\begin{eqnarray} \label{cipc2}
f({\bf \theta}) =\dfrac{1}{2{\pi}\left(\sqrt{\gamma^2+1}-\gamma\cos{\left(\theta-\omega\right)}\right)}
\end{eqnarray}
since $\alpha$ (\ref{a}) may also be expressed as $\alpha=\gamma\cos{\left(\theta-\omega\right)}$, where $\theta$ denotes the angular variable, $\omega$ is the location, $-\pi <\theta, \ \omega < \pi$, and $\gamma \geq 0$ is the concentration parameter. When $\gamma=0$, the distribution reduces to the circular uniform\footnote{Its density function is given by $f\left(\theta \right)=\left(2\pi\right)^{-1}$.}. We will denote the distribution with the density given in (\ref{cipc}) as the circular isotropic projected Cauchy (CIPC) distribution\footnote{We emphasise that the isotropy hereafter refers exclusively to unit variance.}.

Consider now a random variable $X$ on the real line, which we wrap around the circumference of a circle with unit radius by $\theta=X(\mod{2\pi})$. If $X$ follows a Cauchy distribution, then $\theta$ follows a WC distribution on the circle, whose density function is given by \citep[pg.~51]{mardia2000}:
\begin{eqnarray} \label{wc}
f(\theta)=\frac{1-\lambda^2}{2\pi\left[1+\lambda^2-2\lambda\cos\left(\theta-\omega\right)\right]},
\end{eqnarray}
where $\omega$ denotes the location, and $\lambda \in [0, 1)$ acts as the concentration parameter. A close examination reveals that the density in (\ref{cipc2}) is a different parameterisation of the WC distribution in (\ref{wc}), where $\sqrt{\gamma^2+1}=\frac{1+\lambda^2}{1-\lambda^2}$ and $\gamma=\frac{2\lambda}{1-\lambda^2}$ or, conversely, $\lambda=(\sqrt{\gamma^2+1}-1)/\gamma$. This remarkable result shows that the WC distribution can also be obtained by projecting the bivariate Cauchy distribution onto the circle. This appealing equivalence is not observed for the Wrapped Normal and the PN \citep{presnell1998}. The question of whether there are more wrapped distributions that can be created by projection is now raised.

\cite{jones2005} stated that \emph{the WC distribution does not arise from projecting a bivariate spherically symmetric distribution with non-zero mean onto the circle}. According to our previous results, this statement is not entirely accurate. However, there is a relationship between the angular central Gaussian distribution and the WC distribution, as pointed out in \citep[pg.~52]{mardia2000}, if $\theta$ follows the angular central Gaussian distribution then $2\theta$ follows the WC. The fact that we derived the WC via a different route enables us to generalise the WC distribution in a straightforward manner, as will be described in the following section. 

\subsection{The Generalised Circular Projected Cauchy Distribution}
The issue of the CIPC, which is also common with the PN, is that the underlying bivariate Cauchy and bivariate normal distributions, respectively, assume an isotropic covariance matrix. We relax this strict assumption by employing one of the conditions imposed in \cite{paine2018}, that is, $\pmb{\Sigma \mu} = \pmb{\mu}$, but not $|\pmb{\Sigma}|=1$. This condition implies that the one eigenvector $\pmb{\xi}_2$ of $\pmb{\Sigma}$ is the normalised location vector $\pmb{\xi}_2=\left(\mu_1,\mu_2\right)^\top/\gamma$, while the other eigenvector can be defined up to the sign as $\pmb{\xi}_1=\left(-\mu_2, \mu_1\right)^\top/\gamma$ or $\pmb{\xi}_1=\left(\mu_2, -\mu_1\right)^\top/\gamma$. The eigenvalue corresponding to the location vector is equal to 1, while the other eigenvalue is equal to $\rho$, hence $|\pmb{\Sigma}|=\rho > 0$ and the inverse of the scatter matrix is given by
\begin{eqnarray} \label{sinv}
\pmb{\Sigma}^{-1}=\frac{1}{\gamma^2}\left(
\begin{array}{cc}
\mu_1^2+\mu_2^2/\rho & \mu_1\mu_2\left(1-1/\rho\right) \\
\mu_1\mu_2\left(1-1/\rho\right) & \mu_2^2+\mu_1^2/\rho
\end{array}
\right) = \pmb{\xi}_1\pmb{\xi}_1^\top/\rho + \pmb{\xi}_2\pmb{\xi}_2^\top.
\end{eqnarray}

Thus, (\ref{pc}) becomes
\begin{eqnarray} \label{gcpc}
f({\bf y})= \frac{1}{2\pi\rho^{1/2}\left(B\sqrt{\gamma^2+1}-\alpha\sqrt{B}\right)}.
\end{eqnarray}

Utilising (\ref{sinv}) and after some calculations, the density in (\ref{gcpc}) may also be expressed in polar coordinates by
\begin{eqnarray} \label{gcpc2}
f(\theta) &=& \frac{\left(2\pi\rho^{1/2}\right)^{-1}}{\left[\left(\cos^2(\theta-\omega)+\frac{\sin^2(\theta-\omega)}{\rho}\right)\sqrt{\gamma^2+1}-\gamma\cos(\theta-\omega)\sqrt{\cos^2(\theta-\omega)+\frac{\sin^2(\theta-\omega)}{\rho}}\right]} \nonumber \\
&=& \frac{1}{2\pi\rho^{1/2}\left(b\sqrt{\gamma^2+1}-\alpha\sqrt{b}\right)},
\end{eqnarray}
where    
$b = \cos^2(\theta-\omega)+\frac{\sin^2(\theta-\omega)}{\rho}$. Alternatively, (\ref{gcpc2}) may be written as 
\begin{eqnarray} \label{gcpc3} 
f(\theta)=\frac{1}{2\pi\rho^{1/2}\cos^2(\theta-\omega)\sqrt{1 + \frac{\tan^2(\theta-\omega)}{\rho}}\left(\sqrt{\left(\gamma^2+1\right)\left(1+\frac{\tan^2(\theta-\omega)}{\rho}\right)}-\gamma \right)}.  
\end{eqnarray}

We will call the distribution whose density is given by (\ref{gcpc})-(\ref{gcpc3}) the generalised circular projected Cauchy (GCPC) distribution, which is essentially the CPC distribution but with a constrained scatter matrix that also allows for anisotropy. It is important to note that if $\rho=1$, the GCPC distribution reduces to the CIPC distribution (\ref{cipc}). The GCPC distribution exhibits symmetry with respect to $\omega$ such that $f(\theta-\omega)=f(\omega-\theta)$, and is thus symmetrical about $\omega + \pi$. Regrettably, there is no closed form for the cumulative distribution function\footnote{See Appendix for a note on this.}, trigonometric characteristic function, the mean resultant, and its length. This necessitates the application of numerical integration techniques to compute these quantities. The distribution displays bimodality, which can be easily observed by examining the first derivative of the density (\ref{gcpc2})
\begin{small}
\begin{eqnarray*}
& &\frac{\partial\log(f(\theta))}{\partial \omega} = \\
& & \frac{\left(\left(2{\rho}-2\right)\sqrt{{\gamma}^2+1}\cos\left({\theta}-{\omega}\right)\sqrt{\frac{\left({\rho}-1\right)\cos^2\left({\theta}-{\omega}\right)+1}{{\rho}}}+\left(2-2{\rho}\right){\gamma}\cos^2\left({\theta}-{\omega}\right)-{\gamma}\right)\sin\left({\theta}-{\omega}\right)}{\left(\left({\rho}-1\right)\cos^2\left({\theta}-{\omega}\right)+1\right)\left(\sqrt{{\gamma}^2+1}\sqrt{\frac{\left({\rho}-1\right)\cos^2\left({\theta}-{\omega}\right)+1}{{\rho}}}-{\gamma}\cos\left({\theta}-{\omega}\right)\right)}.
\end{eqnarray*}
\end{small}

Setting the above expression to zero, we obtain
\begin{eqnarray*}
& & \left(2{\rho}-2\right)\sqrt{{\gamma}^2+1}\cos\left({\theta}-{\omega}\right)\sqrt{\frac{\left({\rho}-1\right)\cos^2\left({\theta}-{\omega}\right)+1}{{\rho}}}\sin\left({\theta}-{\omega}\right) \\
& & +\left(\left(2-2{\rho}\right){\gamma}\cos^2\left({\theta}-{\omega}\right)-{\gamma}\right)\sin\left({\theta}-{\omega}\right) = 0.
\end{eqnarray*}

By conducting the same computations as those reported in \cite{kato2010}, we observe that the aforementioned equation results in a quartic equation with either four real roots or four complex roots if its discriminant is positive, or two real roots and two complex roots if its discriminant is negative. This finding confirms that the density has at most two modes. Furthermore, the distribution is bimodal when the discriminant is positive and unimodal when the discriminant is negative.

Maximum likelihood estimation of the parameters can be achieved via the Newton-Raphson algorithm; however, the corresponding derivatives are highly complex and extensive. Consequently, we resort to numerical optimisers, such as the simplex algorithm proposed by \cite{nelder1965}, which is accessible in \textit{R} through the \textit{optim()} function. Generating random values from the GCPC distribution is straightforward, requiring only the creation of random vectors from the bivariate Cauchy distribution with a location-constrained scatter matrix and subsequent normalisation by dividing by their norm.

\begin{figure}[!ht]
\centering
\begin{tabular}{cc}
\includegraphics[scale = 0.3]{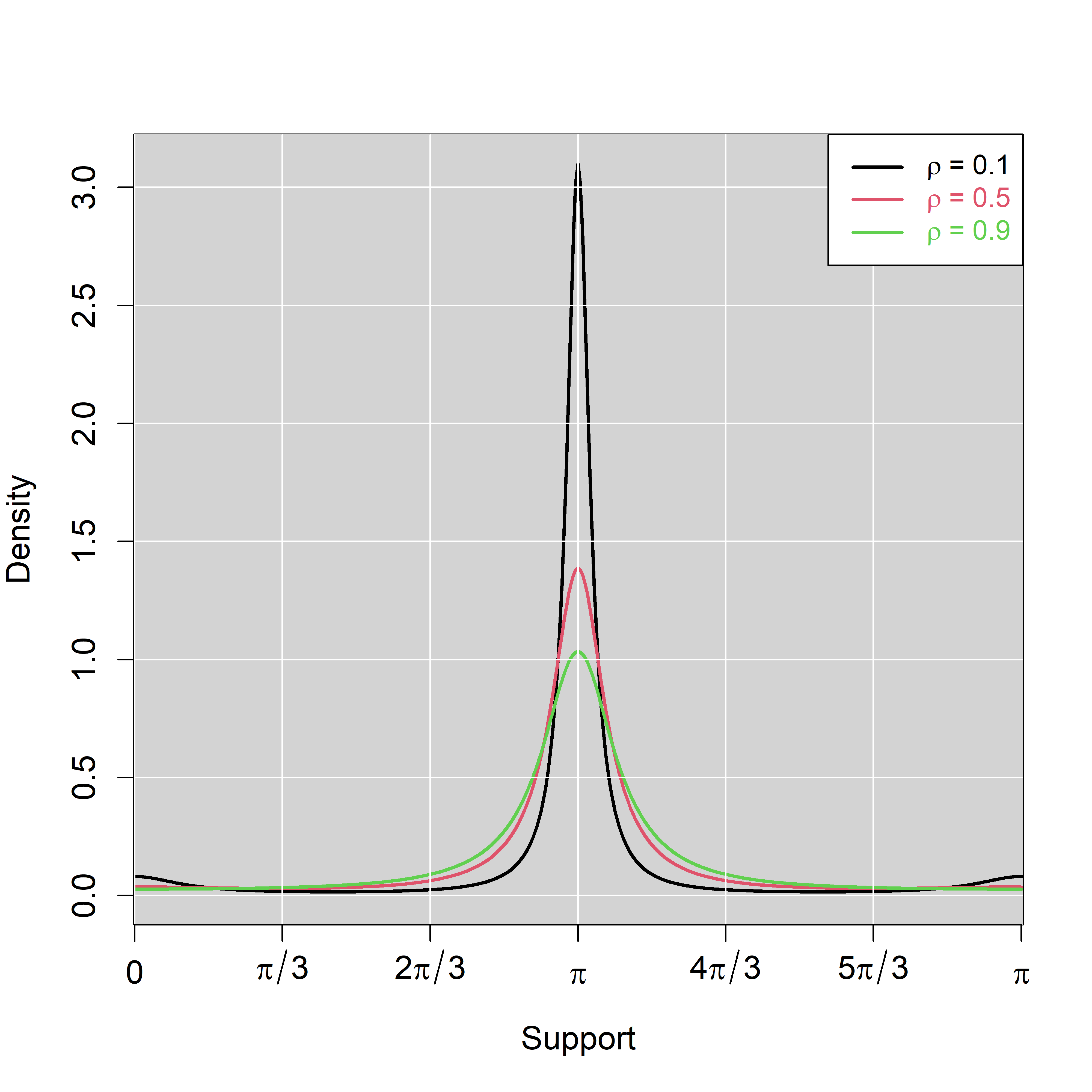}  &
\includegraphics[scale = 0.3]{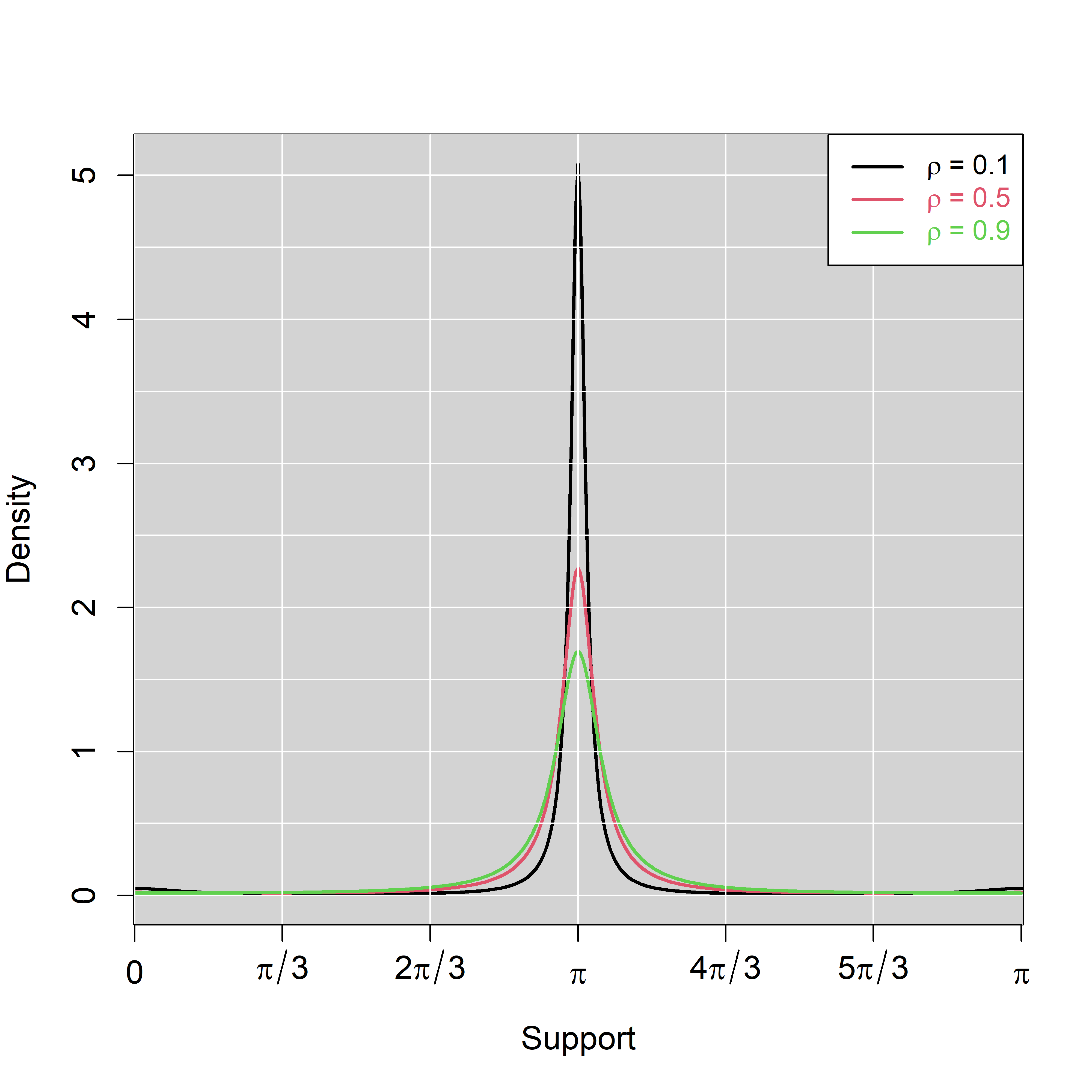}  \\
(a) GCPC distribution with $\gamma=3$  &  (b)  GCPC distribution with $\gamma=5$ \\
\includegraphics[scale = 0.3]{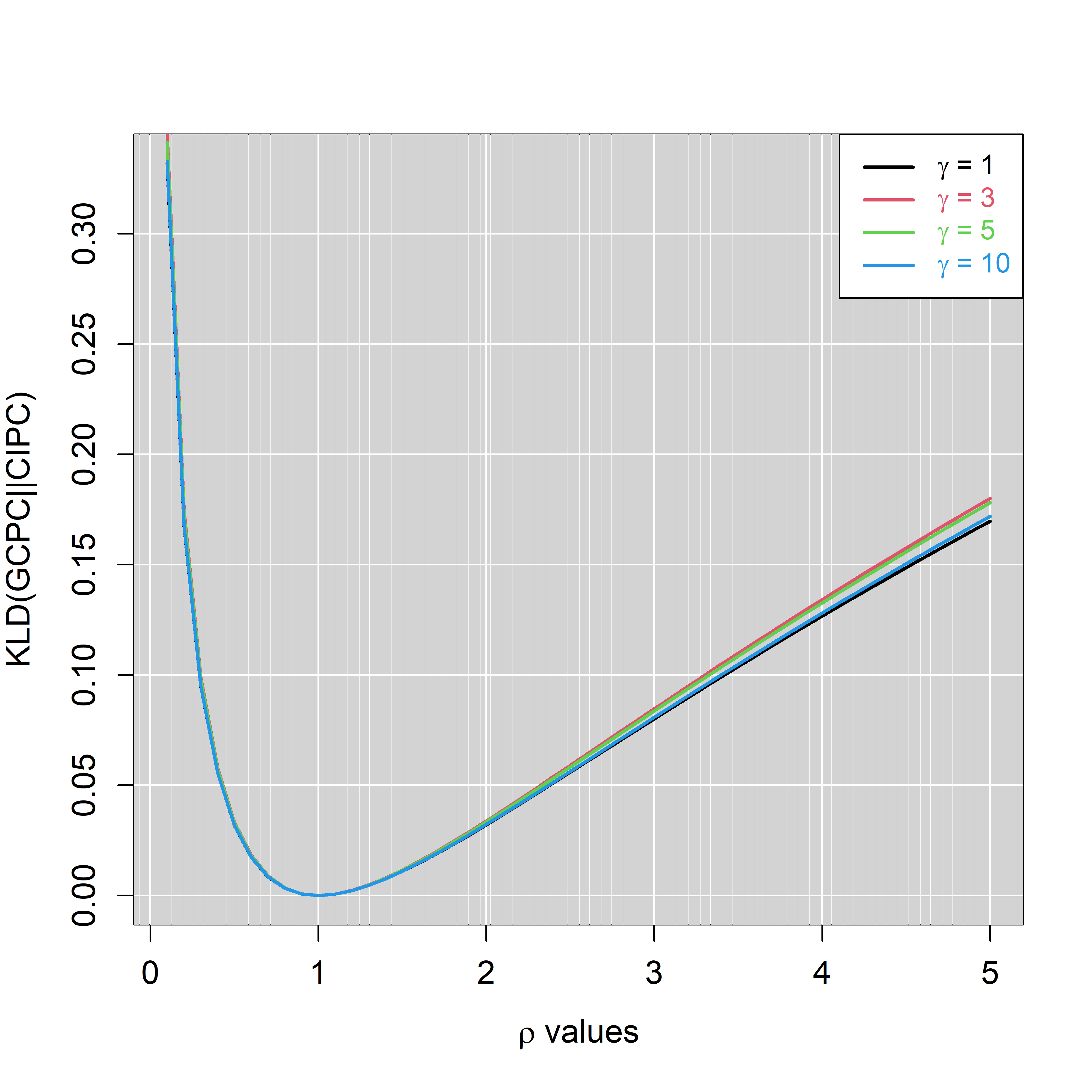}  & \\
(c) KLD of GCPC from the CIPC distribution.  &
\end{tabular}
\caption{Densities of the GCPC distribution when $\omega=\pi$ with (a) $\gamma=3$ and (b) $\gamma=5$, and $\rho=(0.1,0.5,0.9)$. (c) KLD of GCPC from the CIPC distribution.} 
\label{cden1}
\end{figure}

Figures \ref{cden1}(a) and (b) illustrates the GCPC distribution for various values of $\rho$ and $\gamma$. It is worth noting that alternative restrictions or parameterisations, including the addition of an extra parameter (e.g., non-unit variance or a non-zero correlation), could also be developed and utilised. We also computed the Kullback Leibler divergence (KLD) of the GCPC from the CIPC model (KLD(GCPC$\|$CIPC)). We could not find a closed formula for the KLD so we computed it via numerical integration. Figure \ref{cden1}(c) presents the KLD for various values $\rho$ and for different values of $\gamma$. The KLD is not affected by the value of location ($\omega$), but it is affected mainly by the value of $\rho$, and to a lesser degree by the value of $\gamma$, while when $\rho=1$ the KLD is equal to zero as expected. 

\begin{figure}[!ht]
\centering
\label{ckl}
\end{figure}

\subsection{Circular Regression}
The regression setting is straightforward to implement. The response angular data are transformed into their Euclidean coordinates, and the location vector is linked to the covariates in the same way as the spherically projected multivariate linear (SPML) model of \cite{presnell1998}.

\subsubsection{The SPML Regression Model}
The density of the PN is given by
\begin{eqnarray} \label{pn1}
f(\theta)=\frac{1}{2\pi e^{-\gamma^2/2}}\left[1+\frac{\gamma\cos(\theta-\omega)\Phi\left(\gamma\cos(\theta-\omega)\right)}{\phi(\gamma\cos(\theta-\omega))}\right],    
\end{eqnarray}
where $\phi(.)$ and $\Phi(.)$ are the standard normal density and distribution functions,
respectively. Eq. (\ref{pn1}) can also be written as
\begin{eqnarray} \label{pn2}
f\left({\bf y}\right)=\frac{1}{2\pi e^{-\gamma^2/2}}\left[1+\frac{{\bf y}^{\top}\bm{\mu}\Phi\left({\bf y}^{\top}\bm{\mu}\right)}{\phi({\bf y}^{\top}\bm{\mu})}\right],
\end{eqnarray}
where $\bm{\mu} \in \mathbb{R}^2$ denotes the mean vector of the bivariate normal. \cite{presnell1998} linked the mean vector to certain covariates $\bf X$, i.e., $\bm{\mu}_i=\left(\mu_{1i},\mu_{2i}\right)^\top=\left(\bm{\beta}_1^\top x_i,\bm{\beta}_2^\top x_i\right)^\top={\bf B}^\top{\bf x}_i$, where ${\bf B}=\left(\bm{\beta}_1^\top,\bm{\beta}_2^\top\right)$ denotes the matrix of the regression coefficients. The relevant log-likelihood, using (\ref{pn2}), becomes
\begin{eqnarray} \label{pnreg}
\ell_{PN}=-\frac{1}{2}\sum_{i=1}^n\bm{\mu}_i^\top\bm{\mu}_i +\sum_{i=1}^n\log{\left[1+\frac{{\bf y}^\top_i\bm{\mu}_i\Phi\left({\bf y}^\top_i\bm{\mu}_i\right)}{\phi\left({\bf y}^\top_i\bm{\mu}_i\right)}\right]} -n \log\left(2\pi\right).
\end{eqnarray}

Maximisation of (\ref{pnreg}) with respect to the matrix of the regression coefficients ${\bf B}$ is achieved using the Newton-Raphson algorithm or the E-M algorithm \citep{presnell1998}. Note also that $\gamma_i=\|\bm{\mu}_i\|=\left[\left(\bm{\beta}_1^\top x_i\right)^2+\left(\bm{\beta}_2^\top x_i\right)^2\right]^{1/2}$.

\subsubsection{The CIPC and GCPC Regression Models}
The log-likelihood of the CIPC regression model is written as 
\begin{eqnarray} \label{cipcreg}
\ell_{CIPC}=-\sum_{i=1}^n\log{\left(\sqrt{\bm{\mu}_i^\top\bm{\mu}_i+1}-{\bf y}_i^\top\bm{\mu}_i\right)} - n\log{(2\pi)}.
\end{eqnarray}

It is important to note that (\ref{cipcreg}) diverges from the corresponding log-likelihood of the WC distribution (\ref{wc}) in two primary aspects. Firstly, the bivariate representation (Euclidean coordinates) of the angular data is considered, as opposed to their univariate nature. Secondly, the concentration parameter ($\lambda$) of the WC distribution is not maintained as a constant; rather, under the new parametrisation defined in (\ref{cipc}), it is contingent upon $\bm{\mu}_i$. This is akin to the approach employed in the SPML model, as detailed by \cite{presnell1998}. The log-likelihood of the GCPC regression model is written as 

\begin{eqnarray} \label{gcpcreg}
\ell_{GCPC} &=& -\sum_{i=1}^n\log{\left({\bf y}_i^\top\bm{\Sigma}^{-1}_i\left({\bf B}\right){\bf y}_i\sqrt{\bm{\mu}_i^\top\bm{\mu}_i+1}-{\bf y}_i^\top\bm{\mu}_i\sqrt{{\bf y}_i^\top\bm{\Sigma}^{-1}_i\left({\bf B}\right){\bf y}_i}\right)} \nonumber \\
& & -\frac{n}{2}\log(\rho)-n\log{(2\pi)},
\end{eqnarray}
where 
\begin{eqnarray*}
{\bf y}_i^\top\bm{\Sigma}^{-1}_i\left({\bf B}\right){\bf y}_i &=& \bm{\xi}_{1i}\left({\bf B}\right)\bm{\xi}_{1i}\left({\bf B}\right)^\top/\rho+\bm{\xi}_{2i}\left({\bf B}\right)\bm{\xi}_{2i}\left({\bf B}\right)^\top \\
&=& y_{1i}^2\left(\xi_{1i}^2/\rho+\xi_{2i}^2\right)+y_{2i}^2\left(\xi_{2i}^2/\rho+\xi_{1i}^2\right)+2y_{1i}y_{2i}\xi_{1i}\xi_{2i}\left(1-1/\rho\right)
\end{eqnarray*}
and $\xi_{ji}=\frac{\bm{\beta}_j^\top x_i}{\|\bm{\beta}_j^\top x_i\|}=\frac{\mu_{ji}}{\|\bm{\mu}_i\|}=\frac{\mu_{ji}}{\gamma_i}$ for $j=1,2$.

The maximisation of (\ref{cipcreg}) is accomplished through the application of Nelder and Mead's simplex method within the \textit{R} programming environment. In the case of (\ref{gcpcreg}), the same optimisation technique is employed, albeit with a significant distinction. The importance of the initial values necessitates the utilisation of multiple random starts. The presence of $\rho$ adds complexity to the task; to address this issue, the profile log-likelihood of $\rho$ is initially maximised using multiple starting values (e.g., 50 iterations) for the regression coefficients. Subsequently, the resulting optimal values of $\rho$ and the regression coefficients are incorporated as the initial values in a final optimisation procedure.

\section{Spherical Projected Cauchy Distribution} \label{sphere}
The density of the Cauchy distribution in $\mathbb{R}^3$ with some scatter matrix and a location vector $\bm{\mu} \in \mathbb{R}^3$ is given by
\begin{eqnarray*}
f({\bf x}) &=& \frac{\Gamma(2)}{\Gamma(0.5)\pi^{3/2}|\pmb{\Sigma}|^{1/2}}\left[1 + \left({\bf x}-\pmb{\mu}\right)^\top\pmb{\Sigma}^{-1}\left({\bf x}-\pmb{\mu}\right)\right]^{-2} \\
&=& \frac{1}{\pi^2|\pmb{\Sigma}|^{1/2}}\left[1 + \left({\bf x}-\pmb{\mu}\right)^\top\pmb{\Sigma}^{-1}\left({\bf x}-\pmb{\mu}\right)\right]^{-2}.
\end{eqnarray*}

Following the same procedure as before, it turns out that the probability density function of the projected Cauchy variable ${\bf Y} \in \mathbb{S}^2$ is given by
\begin{small}
\begin{eqnarray} \label{spc}
f({\bf y}) &=& \int_0^{\infty}\frac{r^2}{\pi^2|\pmb{\Sigma}|^{1/2}\left[1 + \left(r{\bf y}-\pmb{\mu}\right)^\top\pmb{\Sigma}^{-1}\left({r\bf y}-\pmb{\mu}\right)\right]^2}dr \nonumber \\
&=& \dfrac{B\left(\gamma^2+1\right)\sqrt{\Delta}\left[\operatorname{arctan2}\left(\sqrt{\Delta}, -A\right)-\operatorname{arctan2}\left(\sqrt{\Delta}, A\right)+\pi\right]+2A\Delta}{4\pi^2|\pmb{\Sigma}|^{1/2}B\Delta^2},
\end{eqnarray}
\end{small}
where $\Delta=B\Gamma^2+B-A^2$ (The quantities $A$, $B$ and $\Gamma^2$ were defined in Eq. (\ref{A}), (\ref{B}) and \ref{G2}, respectively) and $\arctan2(y,x)$ is the two-argument arc-tangent:
\begin{eqnarray} \label{atan2}
\arctan2(y, x) = \left\lbrace
\begin{array}{ll}
\arctan(y/x) & \text{if} \ x>0, \\
\arctan(y/x) + \pi & \text{if} \ x<0 \ \text{and} \ y \geq 0, \\
\arctan(y/x) - \pi & \text{if} \ x<0 \ \text{and} \ y<0, \\
\pi/2 & \text{if} \ x=0 \ \text{and} \ y>0, \\
-\pi/2 & \text{if} \ x=0 \ \text{and} \ y<0, \\
\text{undefined} & \text{if} \ x=y=0.
\end{array} \right.
\end{eqnarray}

\subsection{The Spherical Isotropic Projected Cauchy Distribution}
Evidently, when $\pmb{\Sigma}={\bf I}_3$, (\ref{spc}) becomes\footnote{From (\ref{sipc}), we can verify that when $\gamma=0$, which implies that $\pmb{\mu}=(0,0,0)^\top$, then $\alpha=0$ and $\delta=1$, and $\operatorname{arctan2}\left(\sqrt{E}, 0\right)=\pi/2$, hence the numerator is left with $\pi$, whereas the denominator is left with $4\pi^2$. Hence the density function reduces to $\left(4\pi\right)^{-1}$, which is the density function of the spherical uniform distribution.}
\begin{eqnarray} \label{sipc}
f({\bf y})=\dfrac{\left(\gamma^2+1\right)\sqrt{\delta}\left[\operatorname{arctan2}\left(\sqrt{\delta}, -\alpha\right)-\operatorname{arctan2}\left(\sqrt{\delta}, \alpha\right)+\pi\right]+2\alpha\delta}{4\pi^2\delta^2},
\end{eqnarray}
where $\delta=\gamma^2+1-\alpha^2$ (The quantities $\alpha$ and $\gamma$ were defined in Eq. (\ref{a}) and (\ref{g}), respectively). This is the density function of the spherical isotropic projected Cauchy (SIPC) distribution, a rotationally symmetric distribution. 

\subsubsection{The SC Distribution}
A model that appears to be closely related to the SIPC model is the SC distribution \citep{kato2020}, which can be seen as the generalisation of the WC distribution to the sphere (and hypersphere). Its density on $\mathbb{S}^2$ is given by
\begin{eqnarray} \label{sc}
f({\bf y})=\frac{\Gamma\left(1.5\right)}{2\pi^{1.5}}\left(\frac{1-\lambda^2}{1+\lambda^2-2{\bf y}^\top\bm{\mu}}\right)^2,  
\end{eqnarray}
where $\bm{\mu} \in \mathbb{S}^2$ and $\lambda \in [0, 1)$. We stress that, unlike the circular projected Cauchy with an identity covariance matrix, the SIPC is not identical to the SC distribution.

\subsection{The Spherical Elliptically Symmetric Projected Cauchy Distribution}
By imposing the same conditions as in \cite{paine2018}, that is, $\pmb{\Sigma \mu} = \pmb{\mu}$ and $|\pmb{\Sigma}|=1$, (\ref{spc}) becomes
\begin{eqnarray} \label{sespc}
f({\bf y}) = \dfrac{B\left(\gamma^2+1\right)\sqrt{E}\left[\operatorname{arctan2}\left(\sqrt{E}, -\alpha\right)-\operatorname{arctan2}\left(\sqrt{E}, \alpha\right)+\pi\right]+2\alpha E}{4\pi^2BE^2},
\end{eqnarray}
where $E=B\gamma^2+B-\alpha^2$ (The quantities $B$, $\gamma$, and $\alpha$ were defined in Eq. (\ref{B}), \ref{g} and \ref{a}, respectively.). Eq. (\ref{sespc}) defines the density function of the spherical elliptically symmetric projected Cauchy (SESPC) distribution. 

We remind the reader that the scatter matrix $\bm{\Sigma}$ is embedded in (\ref{sespc}) via $A$ (\ref{A}) and $B$ (\ref{B}), terms which are also included in the term $E$. In a similar manner to \cite{paine2018}, the scatter matrix $\bm{\Sigma}$ has a unit eigenvalue, whilst the other two eigenvalues are $0<\rho_1 < \rho_2$, such that $\rho_1\rho_2=1$ and the inverse of $\bm{\Sigma}$ can be written as $\bm{\Sigma}^{-1}=\sum_{j=1}^3\bm{\xi}_j\bm{\xi}_j^\top/\rho_j$, where $\bm{\xi}_1, \bm{\xi}_2$ and $\bm{\xi}_3=\bm{\mu}/\gamma$ is a set of mutually orthogonal unit vectors. Note that $\bm{\xi}_3$ is the location direction, the impact of which is discussed in \S \ \ref{impact}.

As in \cite{paine2018}, we define a pair of unit vectors, $\tilde{\bm{\xi}}_1$ and $\tilde{\bm{\xi}}_2$, which are orthogonal to each other and to the location direction $\bm{\xi}_3$:
$\tilde{\bm{\xi}}_1=\left(-\mu_0^2,\mu_1\mu_2,\mu_1\mu_3\right)^\top/(\gamma\mu_0)$ and $\tilde{\bm{\xi}}_2=\left(0,-\mu_3,\mu_2\right)^\top/\mu_0$, where $\mu_0=(\mu_2^2+\mu_3^2)^{1/2}$. Let us introduce the axes of symmetry $\bm{\xi}_1$ and $\bm{\xi}_2$ to be a rotation of $\tilde{\bm{\xi}}_1$ and $\tilde{\bm{\xi}}_1$ \citep{paine2018}, in the hyperplane orthogonal to $\xi_3=\pmb{\mu}/\gamma$, where the relationship between them is given by 
\begin{eqnarray*}
\bm{\xi}_1 &=& \tilde{\bm{\xi}}_1\cos(\psi)+\tilde{\bm{\xi}}_2\sin(\psi) \ \text{and} \\ 
\bm{\xi}_2 &=& -\tilde{\bm{\xi}}_1\sin(\psi)+\tilde{\bm{\xi}}_2\cos(\psi),
\end{eqnarray*}
where $\psi$ is the angle of rotation. Let us denote $\rho_1=\rho$ and thus $\rho_2=1/\rho$, where $\rho \in (0, 1])$, hence we can write the inverse of the scatter matrix as
\begin{eqnarray*}
\pmb{\Sigma}^{-1} &=& \left[\cos^2(\psi)/\rho+\rho \sin^2(\psi)\right]\tilde{\bm{\xi}}_1\tilde{\bm{\xi}}_1^\top + 
\left[\sin^2(\psi)/\rho+\rho \cos^2(\psi)\right]\tilde{\bm{\xi}}_2\tilde{\bm{\xi}}_2^\top \\
& & +0.5\left(\rho^{-1}-\rho\right)\sin(2\psi)\left(\tilde{\bm{\xi}}_1\tilde{\bm{\xi}}_2^\top + \tilde{\bm{\xi}}_2\tilde{\bm{\xi}}_1^\top\right) + \tilde{\bm{\xi}}_3\tilde{\bm{\xi}}_3^\top.
\end{eqnarray*}
Let us now define $\bm{\theta}=\left(\theta_1,\theta_2\right)^\top$ such that $\theta_1=0.5\left(\rho^{-1}-\rho\right)\cos(2\psi)$ and $\theta_1=0.5\left(\rho^{-1}-\rho\right)\sin(2\psi)$. $\pmb{\Sigma}^{-1}$ then becomes
\begin{eqnarray*}
\pmb{\Sigma}^{-1} &=& {\bf I}_3 + \theta_1\left(\tilde{\bm{\xi}}_1\tilde{\bm{\xi}}_1^\top-\tilde{\bm{\xi}}_2\tilde{\bm{\xi}}_2^\top\right)+\theta_2\left(\tilde{\bm{\xi}}_1\tilde{\bm{\xi}}_2^\top+\tilde{\bm{\xi}}_2\tilde{\bm{\xi}}_1^\top\right) \\
& & +\left[\left(\theta_1^2+\theta_2^2+1\right)^{1/2}-1\right]\left(\tilde{\bm{\xi}}_1\tilde{\bm{\xi}}_1^\top+\tilde{\bm{\xi}}_2\tilde{\bm{\xi}}_2^\top\right).
\end{eqnarray*}

The use of the $\tilde{\bm{\xi}}_i$s axes allows for unconstrained parameter estimation, since, unlike the eigenvalues of $\bm{\Sigma}$, $\theta_1$ and $\theta_2$ lie on the real line. Further, note that the total number of free parameters is five, the same as for the bivariate Cauchy distribution in a tangent space $\mathbb{R}^2$ to the sphere. A remark is that the parameters $\pmb{\mu}$ and $\pmb{\theta}$ are orthogonal, that is, $I_{\bm{\mu},\bm{\theta}}={\bf 0}_{3,2}$, where $I_{\bm{\mu},\bm{\theta}}$ is the Fisher information matrix. 

If $\pmb{\theta}={(0, 0)^\top}$ then $\bm{\Sigma}={\bf I}_3$, and hence the SESPC (\ref{sespc}) reduces to the SIPC (\ref{sipc}), yielding rotational symmetry. The rotational symmetry can hence be tested using the log-likelihood ratio test which, under the null hypothesis, asymptotically follows a $\chi^2$ distribution with two degrees of freedom. Rejection of the rotational symmetry favours the SESPC model (\ref{sespc}) over the SIPC model (\ref{sipc}). Non-parametric bootstrap is an alternative, especially in cases where the sample size is not sufficiently large to allow for the asymptotic null distribution to be valid.  

\subsection{Examining the Influence of the Location Direction as an Eigenvector} \label{impact}
The first condition imposed was that $\bm{\Sigma}\bm{\mu}=\bm{\mu}$, indicating that the second eigenvector (i.e., the eigenvector corresponding to the unit eigenvalue) is equal to the location direction $\bm{\mu}/\gamma$. As \cite{paine2018} assert, this condition enforces symmetry about the eigenvectors of $\pmb{\Sigma}$. Without loss of generality, it is assumed that the eigenvectors are parallel to the coordinate axes; in other words, each element of the vector $\bm{\xi}_j$ is equal to 0 except for the $j$-th element, which is equal to 1. Then, if ${\bf y}=\left(y_1,y_2,y_3\right)^\top$,
\begin{eqnarray} \label{imp}
{\bf y}^\top\bm{\xi}_3=y_3 \ \ \text{and} \ \ {\bf y}^\top\bm{\Sigma}^{-1}{\bf y}=y_3^2+\sum_{j=1}^2y^2_j/\rho_j.
\end{eqnarray}

In this situation, the density in (\ref{sespc}) depends solely on $y_j$ through $y_j^2$ for $j = 1,2$. As a result, the density remains invariant with respect to the sign changes of $y_1, y_2$, that is, $f_{SESPC}(\pm y_1, \pm y_2, y_3) = f_{SESPC}(y_1, y_2, y_3)$, implying reflective symmetry about 0 along the axes defined by $\bm{\xi}_1$ and $\bm{\xi}_2$. This type of symmetry is suggested by ellipse-like contours of constant density inscribed on the sphere, and such contours arise when the density (7) is unimodal. However, a reviewer pointed out that for any given value of $\gamma$, the density will be unimodal for small values of $\rho$ and bimodal for large values of $\rho$, where small/large is relative to the value of $\gamma$. Unfortunately we do not have a mathematical expression for when this case arises and thus, unlike the Kent distribution \cite{kent1982}, we cannot restrict it from happening.

\subsubsection{The ESAG distribution}
The density of the ESAG distribution \citep{paine2018} is given by
\begin{eqnarray} \label{esag}
f\left( {\bf y}; \pmb{\mu}, {\bf V} \right) = \frac{1}{2\pi\left({\bf y}^T{\bf V}^{-1}{\bf y}\right)^{3/2}} 
 \times \exp{\left \lbrace\frac{1}{2}\left[\frac{\left({\bf y}^T\pmb{\mu}\right)^2}{{\bf y}^T{\bf V}^{-1}{\bf y}} - \pmb{\mu}^T\pmb{\mu} \right] \right \rbrace} M_2\left[\frac{\left({\bf y}^T\pmb{\mu}\right)^2}{{\bf y}^T{\bf V}^{-1}{\bf y}}\right],
\end{eqnarray}
where $\bf V$ is used to denote the covariance matrix and $M_2(\alpha) = (1+\alpha^2)\Phi(\alpha) + \alpha\phi(\alpha)$ and $\Phi(.)$, $\phi(.)$ denote the cumulative and probability density function, respectively, of the standard normal distribution.

As mentioned earlier, the parameterisation of the SESPC distribution is the same as that of the ESAG distribution. The main difference between them is that ESAG arises from projecting a multivariate normal on the sphere, whereas the SESPC projects the multivariate Cauchy distribution on the sphere. The $\theta_1$ and $\theta_2$ parameters are termed $\gamma_1$ and $\gamma_2$, respectively, in the ESAG distribution and are defined in exactly the same manner as in the SESPC. When $\gamma_1=\gamma_2=0$, which implies that ${\bf V}={\bf I}_3$, we end up with the PN or the isotropic angular Gaussian (IAG) as termed in \cite{paine2020}

\subsection{Gaphical comparison between SESPC, SIPC, SC and ESAG models}
\begin{figure}[!ht]
\centering
\begin{tabular}{cccc}
\multicolumn{4}{c}{\underline{SESPC distribution}} \\
\includegraphics[scale = 0.35, trim = 20 0 0 0]{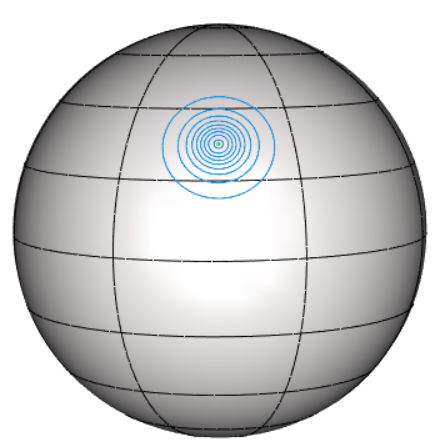}  &
\includegraphics[scale = 0.35, trim = 20 0 0 0]{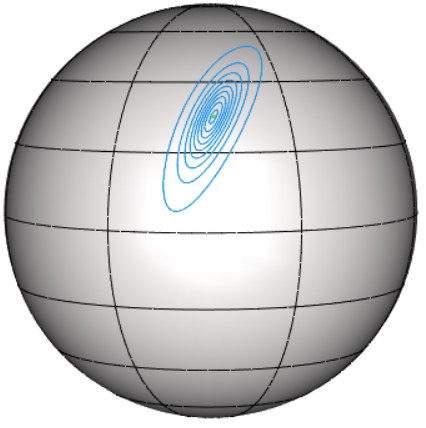}  &
\includegraphics[scale = 0.35, trim = 20 0 0 0]{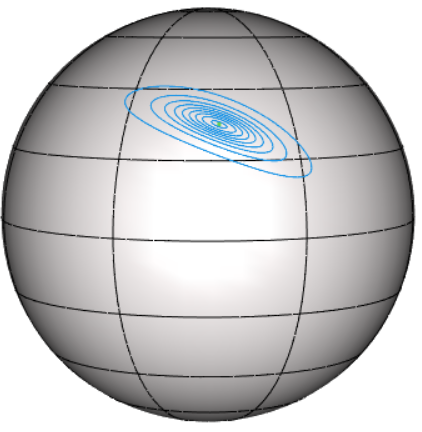}  &
\includegraphics[scale = 0.35, trim = 20 0 0 0]{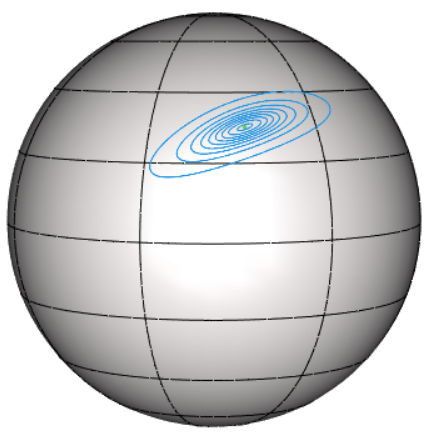}  \\
(a) $\pmb{\theta}=(0, 0)^\top$  &  (b) $\pmb{\theta}=(-1, -1)^\top$ & 
(c) $\pmb{\theta}=(1, 1)^\top$  &  (d) $\pmb{\theta}=(-1, 1)^\top$  \\
&  &  &   \\
\multicolumn{4}{c}{\underline{ESAG distribution}} \\
\includegraphics[scale = 0.35, trim = 20 0 0 0]{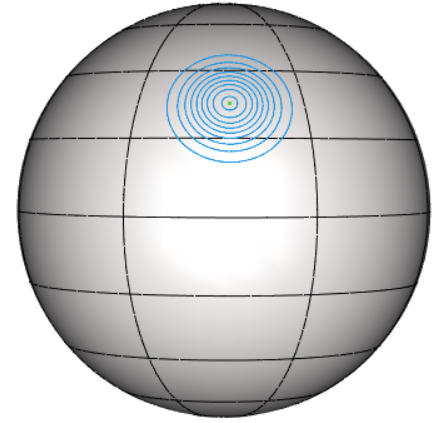}  &
\includegraphics[scale = 0.35, trim = 20 0 0 0]{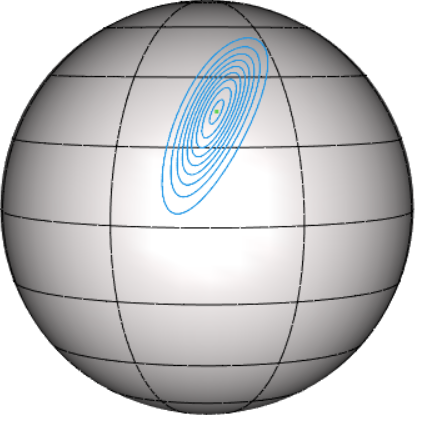}  &
\includegraphics[scale = 0.35, trim = 20 0 0 0]{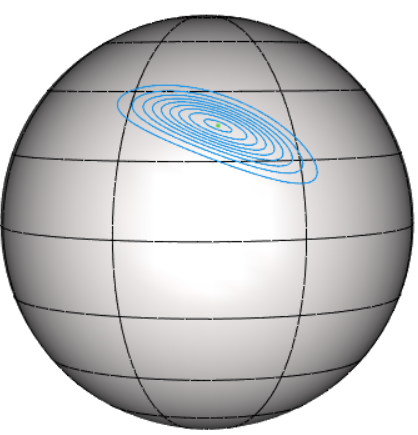}  &
\includegraphics[scale = 0.35, trim = 20 0 0 0]{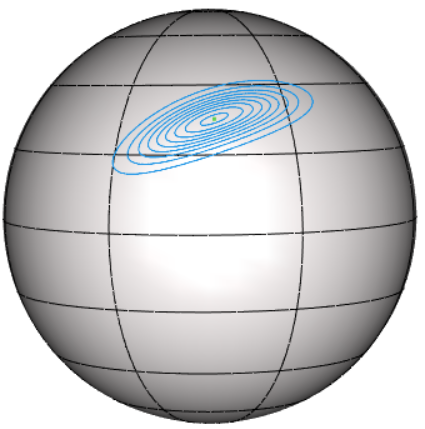}  \\
(e) $\pmb{\gamma}=(0, 0)^\top$  &  (f) $\pmb{\gamma}=(-1, -1)^\top$ & 
(g) $\pmb{\gamma}=(1, 1)^\top$  &  (h) $\pmb{\gamma}=(-1, 1)^\top$  
\end{tabular}
\caption{Contour plots of the SESPC (first row) and ESAG (second row) distributions with $\pmb{\mu}=(5.843,3.057,3.758)^\top$ and various $\pmb{\theta}$ parameter values of the SESPC distribution and $\pmb{\gamma}$ parameter values of the ESAG distribution.}
\label{sden1}
\end{figure}

Figure \ref{sden1} contains some contour plots of the SESPC distribution and of the ESAG distribution for some location vector and various values of $\pmb{\theta}=\left(\theta_1,\theta_2\right)^\top$ (SESPC) and $\pmb{\gamma}=\left(\gamma_1,\gamma_2\right)^\top$ (ESAG). This graphical comparison reveals that the contours of the SESPC distribution appear very similar to those of the ESAG distribution, yet the latter seems to cover more area on the sphere. We further highlight that the contours of the ESAG distribution are very similar to those of the Kent distribution \citep{paine2018}.

Again, we computed the the KLD\footnote{All KLD formulas were computed via numerical integration using the \textit{R} package \textit{SphericalCubature} \citep{nolan2021}.} of the SESPC from the SIPC, of the SIPC from the SC and of the SESPC from the ESAG models. Without loss of generation the location direction was set to $(0,0,\gamma)^\top$, but to allow for a fair comparison between SIPC and SC, when computing KLD(SIPC$\|$SC) the value of $\gamma$ was set equal to 1.

\begin{figure}[!ht]
\centering
\begin{tabular}{cc}
\includegraphics[scale = 0.3]{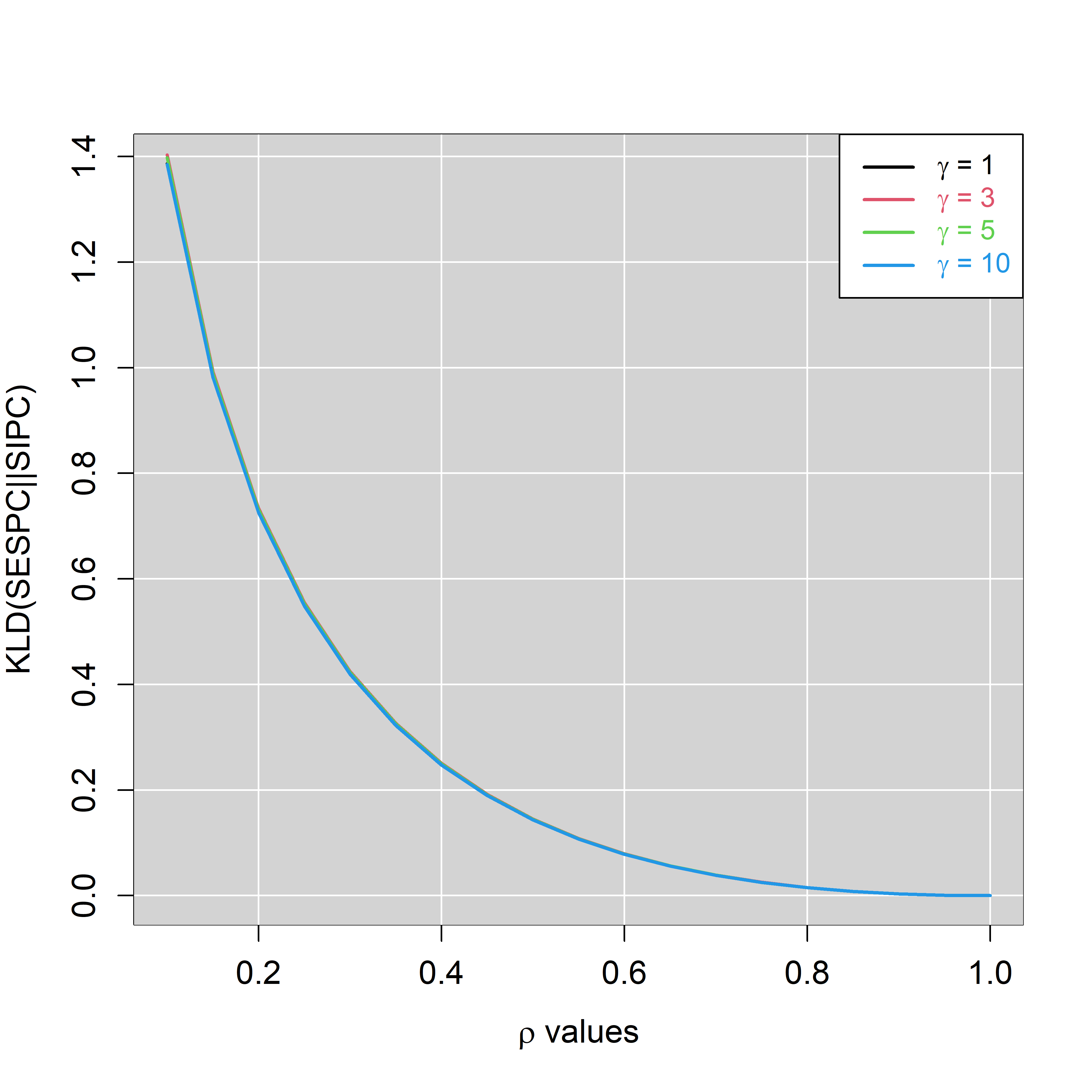} &
\includegraphics[scale = 0.3]{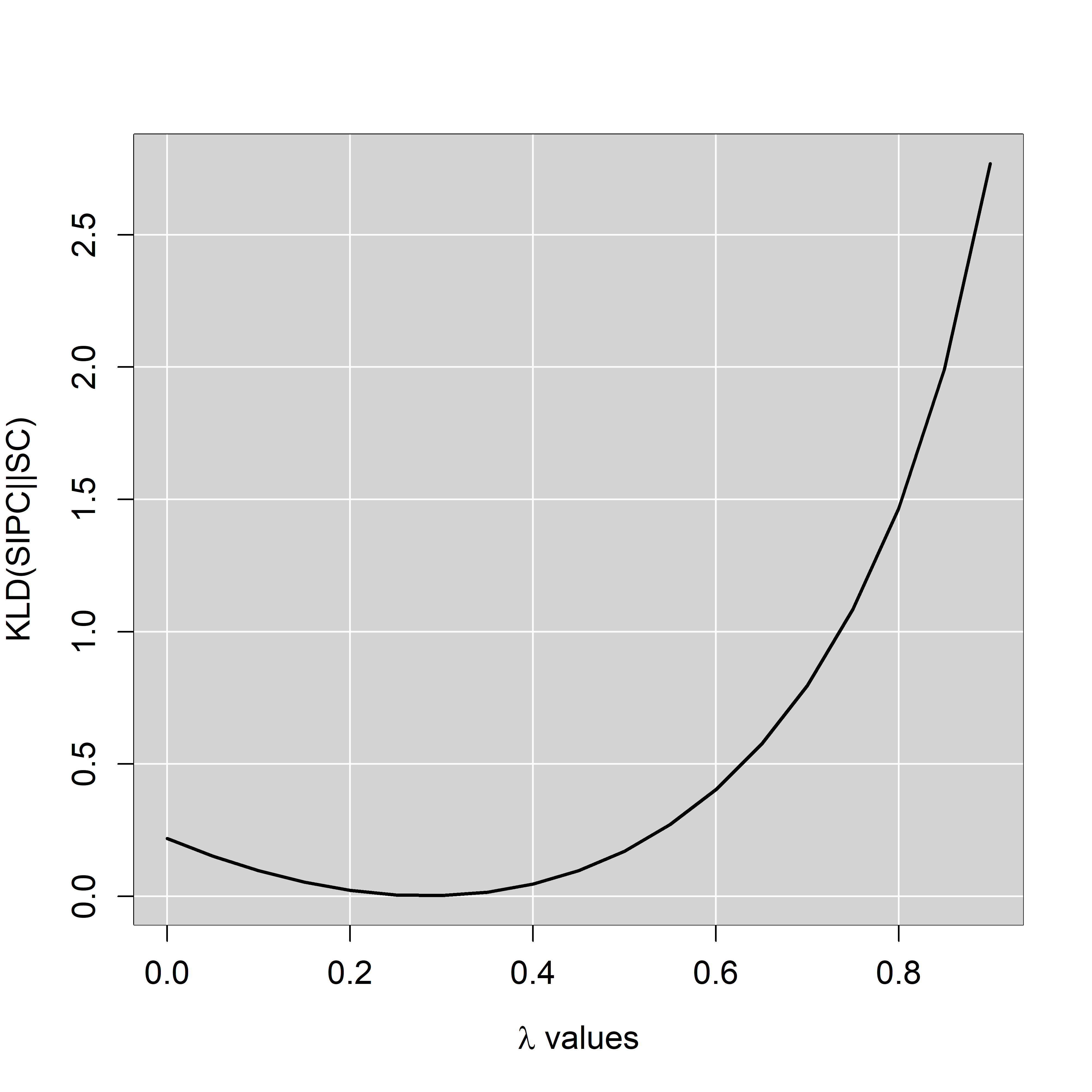} \\
(a) KLD of SESPC from SIPC &  (b) KLD of SIPC from the SC
\end{tabular}
\caption{KLD of SESPC from SIPC and of SIPC from the SC.}
\label{skl}
\end{figure}

Figure \ref{skl}(a) shows the KLD of SESPC from SIPC for different values of $\rho$ and $\gamma$. We remind the reader that $\theta_1=0.5\left(\rho^{-1}-\rho\right)\cos(2\psi)$ and $\theta_1=0.5\left(\rho^{-1}-\rho\right)\sin(2\psi)$. The effect of $\rho$ is depicted in the Figure \ref{skl}(a), and, evidently the effect of the $\gamma$ values is very small compared to the $\rho$ values. The same is true for the $\psi$ values, whose values had a small effect on the KLD, compared to that of the $\rho$ values. Figure \ref{skl}(b) presents the KLD of the SIPC from the SC distribution for different values of the concentration parameter, $\lambda$, of the SC distribution \ref{sc}. Finally, the KLD of the SESPC from the ESAG distribution is slightly affected by the values of $\rho$ and $\psi$, but it is significantly affected by the $\gamma$ values. The mean KLD values are 0.035, 0.588, 1.670 and5.196 for values of $\gamma$ equal to $1,3,5$ and $10$, respectively.

\subsection{Spherical Regression}
In this section, we adopt the second parameterisation (structure 2 of the ESAG distribution, as presented by \cite{paine2020}), to establish a connection between the location direction, $\bm{\mu}$, and covariates, ${\bf X}$. The approach closely resembles that of circular regression, with $\bm{\mu}_i = \left(\bm{\beta}_1^\top{\bf x}_i, \bm{\beta}_2^\top{\bf x}_i, \bm{\beta}_3^\top{\bf x}_i\right)^\top = {\bf B}^\top {\bf x}_i$. The key distinction from circular regression lies in the fact that ${\bf Qy}_i \sim \text{SESPC}\left({\bf B}^\top {\bf x}_i, \bm{\theta}\right)$, where ${\bf Q} \in \textit{O}(3)$ is an estimable rotation matrix. To maximise the associated regression log-likelihood, we optimise with respect to $\bf Q$, $\bf B$, and $\pmb{\theta}$.

To achieve this, a grid search is performed over the orthogonal group, $\textit{O}(3)$. For each $\bf Q$, the response variable ${\bf y}$ is rotated, and the log-likelihood is maximised. The log-likelihood-maximising value of $\bf Q$ is chosen, and the corresponding regression coefficients and $\bm{\theta}$ values are reported.

The log-likelihood ratio test can be employed to test for rotational symmetry (or isotropy) when $\bm{\theta}=(0,0)^\top$, as well as for assessing the significance of one or more covariates. For cases with small sample sizes, the non-parametric bootstrap approach can be considered as an alternative.

\section{Simulation Studies} \label{sims}
The conducted simulation studies comprehensively address the aspects delineated earlier, including an exploration of particular properties pertaining to the maximum likelihood estimation for circular and spherical models, both in the presence and absence of covariates. Furthermore, we draw comparisons between these models and a selection of well-known distributions in the literature. The simulation studies took place in \textit{R} using the package \textit{Directional} \citep{directional2023} and all the relevant functions for the proposed distributions can be found in that package.

\subsection{Comparison of CIPC and GCPC Models}
Initially, we evaluate the bias in the estimated mean vector when the data are generated from the GCPC model with a mean vector equal to $\bm{\mu}=(3,10)^\top$ and $\rho=(0.1,0.3,0.5,0.7,1)$ for various sample sizes, $n=(50, 100, 300, 500, 1000)$. The objectives of this study are: a) to assess the bias of the estimated mean vector between the two models, CIPC and GCPC, and b) to evaluate the power of the log-likelihood ratio in discriminating between the two models. The log-likelihood ratio test statistic in this case follows a mixture of Dirac distribution at 1 and a $\chi^2_1$ distribution with a mixing probability of 0.5\footnote{This is because, under the restriction, we lie at the boundary of the sample space for $\rho$.}.

Table 1 presents the Euclidean distances from the true mean vector of the estimated mean vectors under the CIPC and GCPC models, averaged over 1,000 repetitions. It also provides the estimated power for testing whether $\rho=1$, i.e., distinguishing between the two models. The results clearly demonstrate that when $\rho << 1$, the accuracy of the GCPC model is considerably higher than that of the CIPC model. Furthermore, the discrepancy between them increases with the sample size. As $\rho$ approaches $1$, the differences diminish; when $\rho=0.9$ and $\rho=1$, the average Euclidean distance of the estimated mean vector under the GCPC model is greater than that under the CIPC model. We attribute this phenomenon to the small proportion of times that the log-likelihood ratio rejects the CIPC in favor of the GCPC. To validate this, we computed the averages excluding these cases and observed that the average Euclidean distances were almost identical.  

\begin{table}[ht]
\caption{Average Euclidean distances between the true mean vector $\pmb{\mu}=(3, 10)^\top$ and the estimated mean vector using CIPC and GCPC distributions for various $\rho$ values and sample sizes. The row termed Power contains the estimated power for testing whether $\rho=1$ (when $\rho=1$, the GCPC reduces to the CIPC model).}
\label{tab1}
\centering
\begin{tabular}{l|l|rrrrr}
\hline
            &  & \multicolumn{5}{c}{Sample size} \\  \hline
$\rho$      & Model and Power & n=50 & n=100 & n=300 & n=500 & n=1000 \\  \hline \hline
$\rho=0.1$  & CIPC & 23.467 & 23.043 & 22.824 & 22.505 & 22.470 \\ 
            & GCPC & 9.856 & 6.343 & 2.924 & 2.226 & 1.688 \\ 
            & Power & 0.438 & 0.633 & 0.942 & 0.983 & 0.985 \\  \hline
$\rho=0.3$  &  CIPC & 9.337 & 8.874 & 8.646 & 8.612 & 8.589 \\ 
            & GCPC & 5.762 & 4.475 & 2.655 & 2.026 & 1.355 \\ 
            & Power  & 0.231 & 0.345 & 0.622 & 0.819 & 0.960 \\  \hline
$\rho=0.5$  & CIPC & 5.028 & 4.485 & 4.363 & 4.392 & 4.299 \\ 
            & GCPC & 3.995 & 3.134 & 2.181 & 1.778 & 1.309 \\ 
            & Power & 0.125 & 0.180 & 0.367 & 0.503 & 0.732 \\  \hline
$\rho=0.7$  & CIPC & 3.014 & 2.388 & 2.096 & 2.117 & 2.054 \\ 
            & GCPC & 2.983 & 2.306 & 1.725 & 1.491 & 1.171 \\ 
            & Power & 0.091 & 0.100 & 0.197 & 0.236 & 0.355 \\  \hline
$\rho=0.9$  & CIPC & 1.990 & 1.373 & 0.884 & 0.743 & 0.615 \\ 
            & GCPC & 2.587 & 1.980 & 1.272 & 1.039 & 0.826 \\ 
            & Power & 0.051 & 0.061 & 0.082 & 0.094 & 0.114 \\  \hline
$\rho=1$    & CIPC & 1.761 & 1.207 & 0.682 & 0.543 & 0.378 \\ 
            & GCPC & 2.372 & 1.923 & 1.095 & 0.964 & 0.715 \\ 
            & Power & 0.049 & 0.057 & 0.049 & 0.057 & 0.058 \\  \hline
\end{tabular}
\end{table}

\subsection{Comparison of GCPC, CIPC, and SPML Regression Models}
We examine the discrepancy of the estimated coefficients in a regression setting with a single covariate for the sake of simplicity. The matrix of regression coefficients is defined as
\begin{eqnarray*}
{\bf B}=(\bm{\beta}_1^T,\bm{\beta}_2^T)=
\left(
\begin{array}{cc}
-0.2831 & -0.892 \\
0.066 & 0.090
\end{array}
\right),
\end{eqnarray*}
where the first row corresponds to the constant terms, and the second row represents the slopes. The two columns correspond to $\cos(\theta)$ and $\sin(\theta)$.

The covariate $X$ was generated from a Gamma distribution with shape and rate parameters equal to 2.590 and 0.054, respectively ($Ga(2.590, 0.054)$), and then values of $\theta$ were generated from the GCPC model with a mean vector equal to $\mu_i=\left(\bm{\beta}_1{\bf x}_i,\bm{\beta}_2{\bf x}_i\right)^\top$ and $\rho=(0.1,0.3,0.5,0.7,1)$ for various sample sizes, $n=(50, 100, 300, 500, 1000)$. The objective of this study is to evaluate the discrepancy of the estimated matrix of regression coefficients among the SPML, CIPC, and GCPC regression models, where the discrepancy was calculated using the Frobenius norm, $|\hat{\bf B}-{\bf B}|_F$.

Table \ref{tab2} presents the average Frobenius norm of the difference between the true and estimated regression coefficients using the SPML, CIPC, and GCPC regression models, averaged over 1,000 repetitions. The estimated discrepancy of the SPML model remains constant irrespective of the sample size and the value of $\rho$. In contrast, the GCPC and CIPC models are influenced by the sample size. As the sample size increases, their discrepancy decreases. In the intra-comparison, the discrepancy of the CIPC model is considerably larger than that of the GCPC model for small values of $\rho$. As $\rho$ grows, the GCPC model approaches the CIPC model, and the discrepancies become similar.

\begin{table}[ht]
\caption{Average Frobenius norm between the true and the estimated regression coefficients using SPML, CIPC, and GCPC regression models for various $\rho$ values and sample sizes.}
\label{tab2}
\centering
\begin{tabular}{l|l|rrrrr}
\hline
    &      & \multicolumn{5}{c}{Sample size} \\ \hline \hline
$\rho$     & Model & n=50 & n=100 & n=300 & n=500 & n=1000 \\  \hline
$\rho=0.1$ & SPML & 1.135 & 1.104 & 1.190 & 1.204 & 1.207 \\ 
           & CIPC & 7.053 & 6.497 & 6.211 & 6.189 & 6.162 \\ 
           & CESPC & 2.254 & 1.595 & 1.138 & 1.111 & 0.942 \\  \hline
$\rho=0.3$ & SPML & 1.324 & 1.376 & 1.429 & 1.447 & 1.446 \\ 
           & CIPC & 2.962 & 2.610 & 2.419 & 2.358 & 2.348 \\ 
           & CESPC & 1.374 & 0.977 & 0.547 & 0.458 & 0.318 \\  \hline
$\rho=0.5$ & SPML & 1.455 & 1.516 & 1.558 & 1.560 & 1.574 \\ 
           & CIPC & 1.764 & 1.461 & 1.255 & 1.213 & 1.190 \\ 
           & CESPC & 1.165 & 0.824 & 0.508 & 0.381 & 0.283 \\  \hline
$\rho=0.7$ & SPML & 1.550 & 1.606 & 1.632 & 1.651 & 1.661 \\ 
           & CIPC & 1.292 & 0.914 & 0.659 & 0.612 & 0.574 \\ 
           & CESPC & 1.082 & 0.744 & 0.447 & 0.356 & 0.256 \\  \hline
$\rho=0.9$ & SPML & 1.620 & 1.675 & 1.695 & 1.704 & 1.721 \\ 
           & CIPC & 1.041 & 0.673 & 0.402 & 0.321 & 0.238 \\ 
           & CESPC & 1.003 & 0.657 & 0.405 & 0.322 & 0.233 \\  \hline
$\rho=1.0$ & SPML & 1.634 & 1.687 & 1.729 & 1.736 & 1.741 \\ 
           & CIPC & 1.023 & 0.628 & 0.350 & 0.275 & 0.190 \\ 
           & CESPC & 0.995 & 0.665 & 0.388 & 0.299 & 0.215 \\  \hline
\end{tabular}
\end{table}

\subsection{Comparison of SIPC, SESPC and ESAG Models}
We investigate the bias in the estimated mean vector when the data are generated from the SESPC model with a mean vector equal to $\pmb{\mu}=(5.843,3.057,3.758)^\top$ and $\theta_1=\theta_2=(0.1,0.3,0.5,0.7,1)$ for various sample sizes, $n=(50, 100, 300, 500, 1000)$. The objectives of this study are to: a) evaluate the bias of the estimated mean vector between the SIPC and SESPC models, b) assess the power of the log-likelihood ratio in discriminating between the two models and red{c) compare the bias of the estimated mean vectors, when the data have been generated either from the SESPC or from the ESAG distribution.}

Table \ref{tab3} presents the Euclidean distances of the estimated mean vectors using the SIPC, SESPC and ESAG models, averaged over 1,000 repetitions, when the data were generated from the SESPC distribution. Additionally, the table includes the estimated power for testing whether $\theta_1=\theta_2=0$, i.e., discriminating between the two models. The results clearly demonstrate that as the $\theta$ values increase, the accuracy of the SESPC model becomes significantly higher than that of the SIPC model. Moreover, the discrepancy between the two models increases with the sample size. As expected, as the $\theta$ values approach 0, the differences between the models diminish.
The estimated bias of the mean vectors estimated under the SESPC distribution is significantly less compared to the estimated bias of the mean vectors estimated under the ESAG distribution. This shows that there are practical differences between the two models.

Table \ref{tab3a} presents the Euclidean distances of the estimated mean vectors using the SESPC and ESAG models, averaged over 1,000 repetitions, when the data were generated from the ESAG distribution. The differences between SESPC and ESAG are distinct, but from the opposite side. The estimated bias of the estimated mean vectors under the ESAG distribution is significantly smaller than that of the SESPC distribution, providing again evidence of practical differences between the two models.

\begin{table}[ht]
\caption{Average Euclidean distances between the true mean vector is $\pmb{\mu}=(5.843,3.057,3.758)^\top$ and the estimated mean vector $\hat{\pmb{\mu}}$ using SIPC, SESPC and ESAG distributions. The data have been generated from the SESPC distribution using various values of common $\theta$ parameters. The row termed Power contains the estimated power for testing rotational symmetry ($\theta_1=\theta_2=0$), i.e., the SIPC distribution versus the SESPC distribution.}
\label{tab3}
\centering
\begin{tabular}{l|l|rrrrr}
\hline
&  & \multicolumn{5}{c}{Sample size} \\  \hline
$(\theta_1, \theta_2)$  & Model and Power & n=50 & n=100 & n=300 & n=500 & n=1000 \\    \hline \hline
$\theta_1=\theta_2=0$ & SIPC & 1.106 & 0.739 & 0.422 & 0.332 & 0.221 \\ 
&  SESPC & 1.136 & 0.767 & 0.427 & 0.334 & 0.232 \\ 
&  ESAG & 4.570 & 4.625 & 4.661 & 4.674 & 4.665 \\ 
&  Power & 0.054 & 0.053 & 0.045 & 0.056 & 0.049 \\  \hline
$\theta_1=\theta_2=0.2$ &  SIPC & 1.104 & 0.740 & 0.413 & 0.324 & 0.235 \\ 
&  SESPC & 1.131 & 0.752 & 0.416 & 0.321 & 0.234 \\ 
&  ESAG & 4.574 & 4.631 & 4.656 & 4.657 & 4.665 \\ 
&  Power & 0.265 & 0.484 & 0.912 & 0.996 & 1.000 \\  \hline 
$\theta_1=\theta_2=0.4$ &  SIPC & 1.103 & 0.754 & 0.450 & 0.363 & 0.286 \\ 
&  SESPC & 1.131 & 0.760 & 0.429 & 0.326 & 0.229 \\ 
&  ESAG & 4.593 & 4.644 & 4.652 & 4.658 & 4.665 \\ 
&  Power & 0.745 & 0.974 & 1.000 & 1.000 & 1.000 \\  \hline 
$\theta_1=\theta_2=0.6$ &  SIPC & 1.103 & 0.768 & 0.564 & 0.511 & 0.469 \\ 
&  SESPC & 1.117 & 0.744 & 0.440 & 0.333 & 0.235 \\ 
&  ESAG & 4.653 & 4.623 & 4.651 & 4.661 & 4.668 \\ 
&  Power & 0.971 & 1.000 & 1.000 & 1.000 & 1.000 \\  \hline 
$\theta_1=\theta_2=0.8$ &  SIPC & 1.190 & 0.905 & 0.733 & 0.732 & 0.719 \\ 
&  SESPC & 1.132 & 0.804 & 0.424 & 0.343 & 0.227 \\ 
&  ESAG & 4.650 & 4.641 & 4.658 & 4.666 & 4.671 \\ 
&  Power & 0.999 & 1.000 & 1.000 & 1.000 & 1.000 \\  \hline 
$\theta_1=\theta_2=1.0$ &  SIPC & 1.244 & 1.041 & 0.976 & 0.968 & 0.986 \\ 
&  SESPC & 1.127 & 0.766 & 0.432 & 0.334 & 0.239 \\ 
&  ESAG & 4.675 & 4.640 & 4.667 & 4.663 & 4.675 \\ 
&  Power & 1.000 & 1.000 & 1.000 & 1.000 & 1.000 \\  \hline 
\end{tabular}
\end{table}

\begin{table}[ht]
\caption{Average Euclidean distances between the true mean vector is $\pmb{\mu}=(5.843,3.057,3.758)^\top$ and the estimated mean vector $\hat{\pmb{\mu}}$ using SESPC and ESAG distributions. The data have been generated from the ESAG distribution using various values of common $\gamma$ parameters.}
\label{tab3a}
\centering
\begin{tabular}{l|l|rrrrr}
\hline
&  & \multicolumn{5}{c}{Sample size} \\  \hline
$(\gamma_1, \gamma_2)$  & Model and Power & n=50 & n=100 & n=300 & n=500 & n=1000 \\    \hline \hline
$\gamma_1=\gamma_2=0$ & SESPC & 3.636 & 3.389 & 3.199 & 3.175 & 3.160 \\ 
&  ESAG & 0.540 & 0.371 & 0.206 & 0.158 & 0.114 \\  \hline
$\gamma_1=\gamma_2=0.2$ &  SESPC & 3.609 & 3.406 & 3.205 & 3.178 & 3.154 \\ 
&  ESAG & 0.538 & 0.377 & 0.204 & 0.161 & 0.117 \\  \hline 
$\gamma_1=\gamma_2=0.4$ &  SESPC & 3.660 & 3.317 & 3.223 & 3.152 & 3.142 \\ 
&  ESAG & 0.551 & 0.368 & 0.212 & 0.161 & 0.115 \\  \hline 
$\gamma_1=\gamma_2=0.6$ &  SESPC & 3.729 & 3.396 & 3.191 & 3.181 & 3.135 \\ 
&  ESAG & 0.550 & 0.376 & 0.211 & 0.164 & 0.114 \\  \hline 
$\gamma_1=\gamma_2=0.8$ &  SESPC & 3.705 & 3.418 & 3.205 & 3.168 & 3.152 \\ 
&  ESAG & 0.551 & 0.378 & 0.210 & 0.159 & 0.115 \\  \hline 
$\gamma_1=\gamma_2=1.0$ &  SESPC & 3.706 & 3.409 & 3.220 & 3.178 & 3.148 \\ 
&  ESAG & 0.565 & 0.398 & 0.211 & 0.166 & 0.119 \\  \hline 
\end{tabular}
\end{table}

\subsection{Comparative Analysis of SESPC, SIPC, SC and ESAG Models}
We evaluate the accuracy of the SESPC, SIPC, SC and ESAG models in estimating the mean direction of the fitted model, ${\bf m}={\pmb{\mu}}/\gamma$, under the conditions where the data are generated using the SESPC model for a range of $\theta_1$ and $\theta_2$ values, and when the data are generated using the ESAG model under the same range of $\gamma_1$ and $\gamma_2$ values. The measure employed for this analysis is $\sqrt{2\left[1-\text{E}\left(\widehat{\bf m}^\top{\bf m}\right)\right]}$, with the expectation approximated by Monte Carlo.

Table \ref{tab4} presents the following quantity:
\begin{eqnarray*}
\text{Error}\left(\widehat{\bf m}\right)=\sqrt{2\left[1-B^{-1}\sum_{b=1}^B\left(\widehat{\bf m}_i^\top{\bf m}\right)\right]},
\end{eqnarray*}
where $\left(\widehat{\bf m}_i\right)$ represents the estimated mean direction of the fitted model for the $i$-run out of $B$ Monte Carlo runs. 

When the data have been generated from the SESPC distribution (Table \ref{tab4}), regardless of the true model values, the errors for all distributions are very similar for large sample sizes ($n \geq 300$), when the true mean vector is a unit vector, i.e. $\gamma=1$. However, when the true mean vector does not lie on the sphere, our simulation studies (not presented here) have showed that there are differences between the two major competing models (SESPC and ESAG).

\begin{table}[ht]
\caption{The $\text{Error}\left(\widehat{\bf m}\right)$ of the fitted models when the true mean vector is $\pmb{\mu}=(5.843,3.057,3.758)^\top$ and hence the true mean direction is ${\bf m}=(0.770,0.403,0.495)^\top$, when the data are generated from the SESPC distribution with mean equal to $\bf m$ (in order to allow for a fair comparison with the SC distribution) and various values of common $\theta$ (or $\gamma$ for the case of the ESAG) parameters.}
\label{tab4}
\centering
\begin{tabular}{l|l|rrrrr|rrrrr}
\hline
&  & \multicolumn{5}{c}{SESPC is the true model} & \multicolumn{5}{c}{ESAG is the true model} \\  \hline
$(\theta_1, \theta_2)$  & Model & n=50 & n=100 & n=300 & n=500 & n=1000 & n=50 & n=100 & n=300 & n=500 & n=1000\\    \hline \hline
$\theta_1=\theta_2=0$ &  SC & 0.271 & 0.198 & 0.111 & 0.087 & 0.061 & 0.214 & 0.156 & 0.089 & 0.070 & 0.049 \\ 
&  SIPC & 0.270 & 0.195 & 0.109 & 0.086 & 0.060 & 0.223 & 0.162 & 0.093 & 0.074 & 0.051 \\ 
&  SESPC & 0.293 & 0.200 & 0.110 & 0.087 & 0.060 & 0.222 & 0.162 & 0.093 & 0.074 & 0.051 \\ 
&  ESAG & 0.349 & 0.230 & 0.120 & 0.092 & 0.064 & 0.216 & 0.156 & 0.088 & 0.069 & 0.048 \\  \hline 
$\theta_1=\theta_2=0.2$ &  SC & 0.284 & 0.207 & 0.114 & 0.088 & 0.062 & 0.224 & 0.162 & 0.091 & 0.073 & 0.051 \\ 
&  SIPC & 0.285 & 0.209 & 0.115 & 0.089 & 0.062 & 0.238 & 0.174 & 0.097 & 0.077 & 0.055 \\ 
&  SESPC & 0.305 & 0.204 & 0.111 & 0.086 & 0.059 & 0.227 & 0.162 & 0.092 & 0.073 & 0.051 \\ 
&  ESAG & 0.358 & 0.224 & 0.129 & 0.093 & 0.064 & 0.224 & 0.150 & 0.087 & 0.068 & 0.047 \\  \hline 
$\theta_1=\theta_2=0.4$ &  SC & 0.293 & 0.218 & 0.125 & 0.095 & 0.068 & 0.248 & 0.175 & 0.101 & 0.077 & 0.055 \\ 
&  SIPC & 0.310 & 0.230 & 0.132 & 0.101 & 0.072 & 0.274 & 0.194 & 0.113 & 0.086 & 0.062 \\ 
&  SESPC & 0.285 & 0.206 & 0.109 & 0.085 & 0.060 & 0.226 & 0.155 & 0.089 & 0.068 & 0.048 \\ 
&  ESAG & 0.319 & 0.222 & 0.116 & 0.089 & 0.064 & 0.209 & 0.144 & 0.095 & 0.064 & 0.045 \\  \hline 
$\theta_1=\theta_2=0.6$ &  SC & 0.335 & 0.247 & 0.140 & 0.105 & 0.078 & 0.266 & 0.188 & 0.113 & 0.085 & 0.062 \\ 
&  SIPC & 0.365 & 0.275 & 0.157 & 0.117 & 0.088 & 0.309 & 0.219 & 0.132 & 0.101 & 0.073 \\ 
&  SESPC & 0.272 & 0.183 & 0.103 & 0.100 & 0.055 & 0.206 & 0.141 & 0.080 & 0.061 & 0.044 \\ 
&  ESAG & 0.300 & 0.206 & 0.122 & 0.097 & 0.056 & 0.194 & 0.133 & 0.075 & 0.058 & 0.076 \\  \hline 
$\theta_1=\theta_2=0.8$ &  SC & 0.357 & 0.257 & 0.151 & 0.119 & 0.087 & 0.291 & 0.211 & 0.121 & 0.097 & 0.070 \\ 
&  SIPC & 0.410 & 0.301 & 0.181 & 0.144 & 0.105 & 0.357 & 0.259 & 0.153 & 0.123 & 0.088 \\ 
&  SESPC & 0.238 & 0.158 & 0.091 & 0.069 & 0.102 & 0.181 & 0.129 & 0.069 & 0.083 & 0.039 \\ 
&  ESAG & 0.252 & 0.163 & 0.110 & 0.070 & 0.102 & 0.169 & 0.120 & 0.067 & 0.051 & 0.073 \\  \hline 
$\theta_1=\theta_2=1.0$ &  SC & 0.396 & 0.278 & 0.172 & 0.130 & 0.095 & 0.330 & 0.230 & 0.143 & 0.108 & 0.076 \\ 
&  SIPC & 0.471 & 0.340 & 0.221 & 0.167 & 0.123 & 0.414 & 0.299 & 0.192 & 0.146 & 0.103 \\ 
&  SESPC & 0.213 & 0.142 & 0.135 & 0.088 & 0.099 & 0.183 & 0.112 & 0.064 & 0.050 & 0.096 \\ 
&  ESAG & 0.246 & 0.155 & 0.120 & 0.141 & 0.100 & 0.160 & 0.107 & 0.062 & 0.080 & 0.114 \\  \hline
\end{tabular}
\end{table}

\subsection{Evaluation of SIPC, SESPC and ESAG Regression Models}
We generated data from a single-covariate regression model, with the matrix of coefficients being defined as:
\begin{eqnarray*}
{\bf B}=
\left(
\begin{array}{ccc}
-1 & 1 & -0.5 \\
0.4 & -0.5 & 0.3
\end{array}
\right),
\end{eqnarray*}
where the first row represents the constant term, and the second row corresponds to the slope. The data were generated from the SESPC model with $\theta_1=\theta_2=0$ (corresponding to the SIPC model) and with $\theta_1=-2$ and $\theta_2=2$ for various sample sizes. We repeated this scenario when the data were generated from the ESAG distribution. We then estimated the regression model using both SIPC, SESPC and ESAG (without applying the rotation matrix $\bf Q$) and computed the Frobenius norm between the true and estimated regression coefficient matrices. Table \ref{tab5} displays the average norms over 1,000 iterations.

Under the SIPC model, the results between the SIPC and SESPC regression models are comparable, and they both outperform ESAG regression model. However, when the $\pmb{\theta}$ parameters are non-zero, differences in the estimated discrepancy of the regression coefficients become apparent and SESPC outperforms SIPC, and both outperform the ESAG regression model. On the contrary, when the data are generated using the ESAG distribution with isotropic covariance matrix, the ESAG regression model clearly outperforms the suggested regression models. When the $\pmb{\gamma}$ parameters of the ESAG distribution are non-zero, ESAG outperforms, as expected, the suggested regression models, with the exception of large sample sizes, where the performance of the ESAG is slightly worse than that of the SIPC regression model.

\begin{table}[ht]
\caption{Average Frobenius norm between the true and the estimated regression coefficients using SIPC, SESPC and ESAG regression models for various sample sizes when the data have been generated using the SESPC and the ESAG distributions.}
\label{tab5}
\centering
\begin{tabular}{l|l|rrrrr}
\hline
True model                 &       & \multicolumn{5}{c}{Sample size} \\ \hline 
SESPC                      & Model & n=50  & n=100 & n=300 & n=500 & n=1000 \\  \hline \hline
$\theta_1=\theta_2=0$      & SIPC  & 0.787 & 0.519 & 0.283 & 0.215 & 0.152  \\ 
                           & SESPC & 0.887 & 0.580 & 0.296 & 0.222 & 0.156  \\ 
                           & ESAG  & 0.858 & 0.692 & 0.560 & 0.538 & 0.525  \\  \hline
$\theta_1=-2, \theta_2=2$  & SIPC  & 1.181 & 0.854 & 0.615 & 0.558 & 0.521  \\ 
                           & SESPC & 0.870 & 0.730 & 0.559 & 0.521 & 0.488  \\ 
                           & ESAG  & 0.915 & 0.860 & 0.882 & 0.885 & 0.859  \\  \hline
ESAG                       & Model & n=50  & n=100 & n=300 & n=500 & n=1000 \\  \hline \hline 
$\gamma_1=\gamma_2=0$      & SIPC  & 1.019 & 0.762 & 0.584 & 0.540 & 0.505  \\ 
`                          & SESPC & 1.008 & 0.755 & 0.582 & 0.541 & 0.506  \\ 
                           & ESAG  & 0.735 & 0.481 & 0.240 & 0.179 & 0.118  \\   \hline
$\gamma_1=-2, \gamma_2=2$  & SIPC  & 1.216 & 0.825 & 0.452 & 0.354 & 0.261  \\ 
                           & SESPC & 0.952 & 0.802 & 0.695 & 0.681 & 0.654  \\ 
                           & ESAG  & 0.525 & 0.392 & 0.315 & 0.306 & 0.279  \\  \hline   
\end{tabular}
\end{table}

\section{Empirical Illustrations with Real Data: A Comparative Analysis} \label{real}
In this section, we present a series of real-world data examples to demonstrate the effectiveness and advantages of our proposed models. Through these empirical illustrations, we aim to provide a comparative analysis that highlights the superior performance of the suggested models over some of the prevalent models in the field. This comprehensive comparison serves to emphasise the practical applicability and relevance of our newly developed models in addressing real-world problems and challenges. 

\subsection{Circular Data without Covariates}
\begin{figure}[!ht]
\centering
\includegraphics[scale = 0.5]{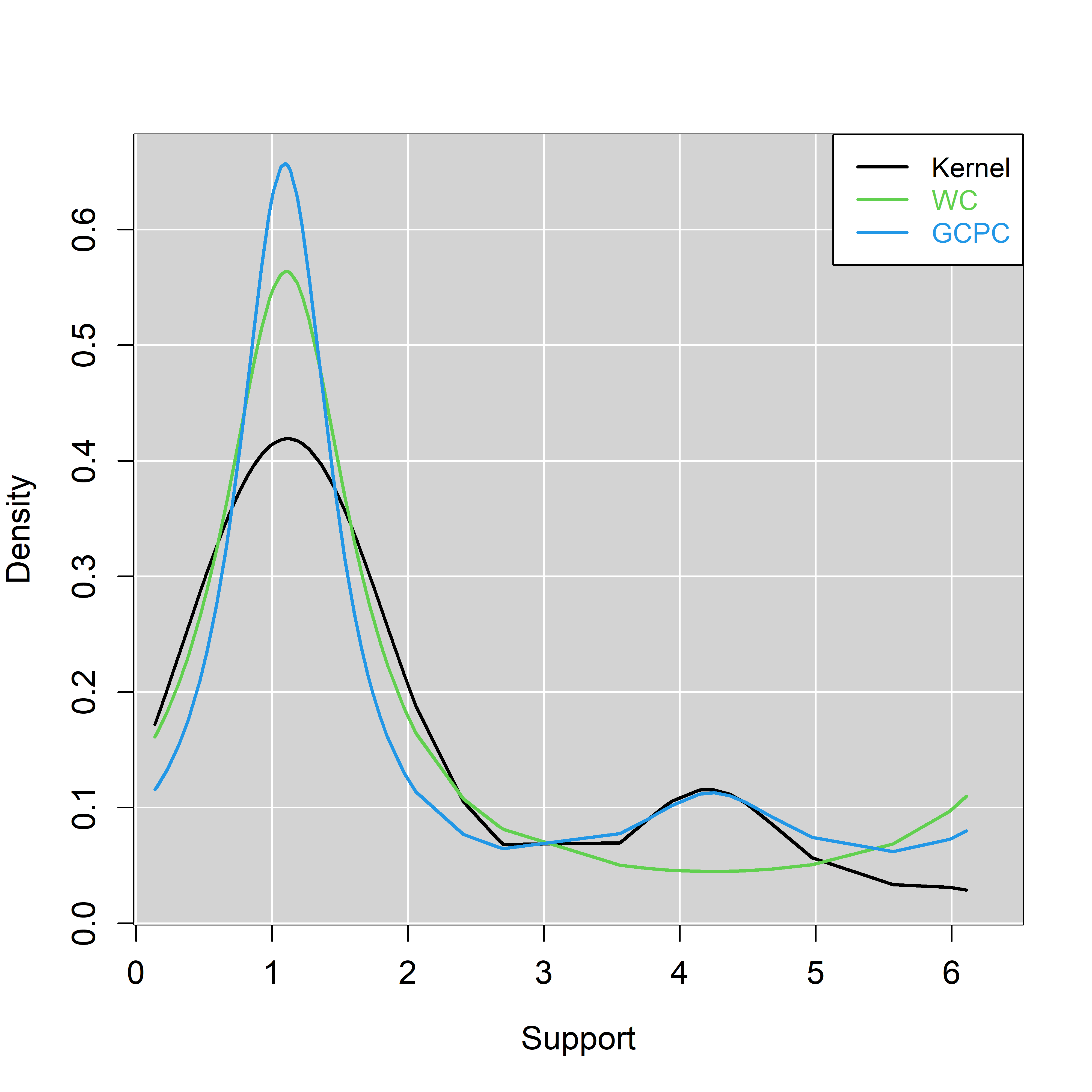} 
\caption{The kernel density estimate, and the WC \& GCPC densities of the turtles dataset.}
\label{fisherb3}
\end{figure}

The first example pertains to a dataset that comprises measurements of the directions taken by 76 turtles after undergoing a specific treatment\footnote{This dataset can be accessed and downloaded from the \textit{R} package \textit{circular} \citep{circular2017} under the name \textit{fisherB3} or \textit{fisherB3c}.} \citep[pg.~241]{fisher1995}. We present the estimated parameters of the CIPC and GCPC models in Table \ref{est}. In general, the parameters appear to be quite similar, except for the $\gamma$ parameter for the CIPC model, which exhibits a higher value. The p-value for the hypothesis that $\rho=1$ is equal to 0.0002, indicating a clear preference for the GCPC model to the CIPC model. Figure \ref{fisherb3} displays the kernel density estimate of the data alongside the fitted densities of the CIPC and the GCPC models. Both the CIPC and GCPC models produce a higher density for the first mode; however, only the GCPC model is successful in capturing the second mode. This result emphasises the superior performance of the GCPC model in this particular example.

\begin{table}[ht]
\caption{Estimated parameters for the CIPC and GCPC models for the Turtles dataset.}
\label{est}
\centering
\begin{tabular}{l|cccccc}
\hline
Model &$\hat{\mu}$  &  $\hat{\bm{\mu}}$ in $\hat{\mathbb{S}^1}$  &  $\hat{\bm{\mu}}$ in $\mathbb{R}^2$  &  $\hat{\gamma}$ &  $\hat{\rho}$  &  Log-likelihood \\  \hline \hline 
CIPC  & 1.107  &  $(0.448, 0.894)^\top$  &  $(0.730, 1.458)^\top$ &  1.630  &       & -113.248 \\
GCPC & 1.094  &  $(0.459, 0.889)^\top$  &  $(0.458, 0.888)^\top$ &  0.999  & 0.341 & -109.857 \\ \hline
\end{tabular}
\end{table}

\subsection{Circular Data with Covariates}
The second example explores the regression setting using the \textit{wind} data\footnote{The dataset is the \textit{speed.wind2} accessible via the \textit{R} package \textit{NPCirc} \citep{npcirc2014}.}. The dataset contains hourly observations of wind direction and wind speed during the winter season (from November to February) from 2003 until 2012 on the Atlantic coast of Galicia (NW–Spain). Data are recorded by a buoy located at longitude -0.21$^{\circ}$ east and latitude 43.5$^{\circ}$ north in the Atlantic Ocean. However, due to the time series nature of the data, only a subset of 200 points was analysed. This dataset, as analysed in \cite{oliveira2014}, was obtained by selecting observations with a lag period of 95 hours.

Table \ref{creg_res} presents the estimated regression parameters using the PN, CIPC, and GCPC distributions for the fitted regression models. The log-likelihood values for the CIPC and GCPC regression models were found to be $-363.370$ and $-334.732$, respectively. Furthermore, the $\rho$ parameter of the GCPC distribution was estimated to be 0.202 with a standard error of 0.039. The circular correlation coefficient \citep[pg.~178]{jammalamadaka2001} between the observed and fitted angular data is equal to $0.044, 0.042$, and $0.080$ for the SPML, CIPC, and GCPC regression models, respectively.

\begin{table}[ht]
\caption{Estimated regression parameters (their standard errors appear within parentheses) for the PN, CIPC, and GCPC regression models fitted to the wind data.}
\label{creg_res}
\centering
\begin{tabular}{l|cc|cc|cc} \hline
Model            & \multicolumn{2}{c}{PN} & \multicolumn{2}{c}{CIPC} & \multicolumn{2}{c}{GCPC}   \\ \hline \hline
&  $\cos(y)$  & $\sin(y)$ & $\cos(y)$ & $\sin(y)$ & $\cos(y)$ & $\sin(y)$  \\ \hline
$\hat{\bm{\alpha}}$ & 0.333(0.185)  & 0.120(0.181)  & 0.458(0.282)  & 0.209(0.245)  & -0.042 (0.0013) & -0.045(0.0023) \\
$\hat{\bm{\beta}}$  & -0.024(0.021) & -0.007(0.020) & -0.033(0.032) & -0.011(0.028) &  0.009(0.0005) & 0.010(0.0009)  \\ \hline \hline
& Loglik   & $\hat{\rho}$ & Loglik   & $\hat{\rho}$ & Loglik   & $\hat{\rho}$ \\ \hline
& -363.893 &               - & -363.370 & -               & -331.846 & 0.203(0.039)    \\ \hline 
\end{tabular}
\end{table}

\subsection{Spherical Data Without Covariates}
\begin{figure}[!ht]
\centering
\includegraphics[scale = 0.6, trim = 40 0 0 0]{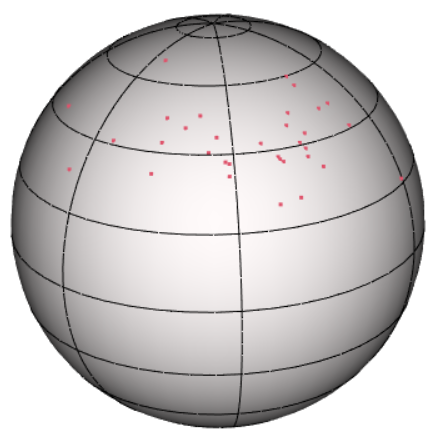} 
\caption{Sphere plot of the Paleomagnetic dataset.}
\label{wood}
\end{figure}

We visually compared the density values of the SC, SIPC, SESPC and ESAG distributions using the \textit{Paleomagnetic pole} dataset \citep{schmidt1976}. This dataset comprises estimates of the Earth's historical magnetic pole position calculated from 33 distinct sites in Tasmania. The data are illustrated in Figure \ref{wood}. Notably, rotational symmetry is rejected (p-value = 0.0026), supporting the findings of \cite{paine2018}.

\begin{figure}[!ht]
\centering
\begin{tabular}{cc}
\includegraphics[scale = 0.55, trim = 0 0 0 0]{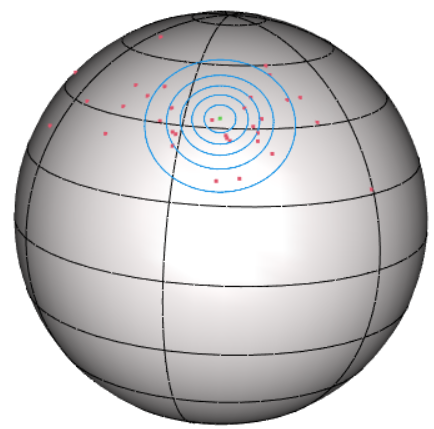} &
\includegraphics[scale = 0.55, trim = 0 0 0 0]{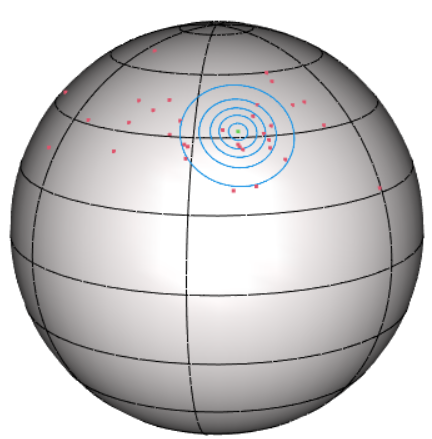} \\
(a) SC &  (b) SIPC \\
\includegraphics[scale = 0.55, trim = 0 0 0 0]{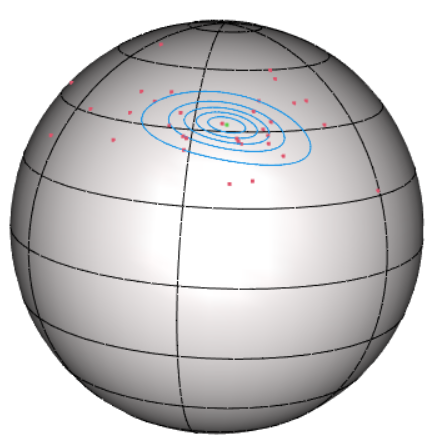}  &
\includegraphics[scale = 0.55, trim = 0 0 0 0]{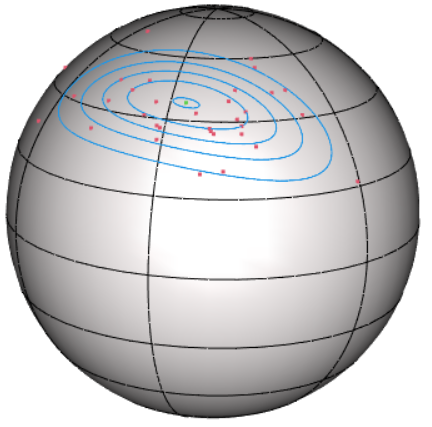}  \\
(c) SESPC & (d) ESAG    
\end{tabular}
\caption{Comparison of the four distributions fitted to the \textit{Paleomagnetic pole} dataset: spherical contour plots of each fitted distribution.}
\label{cden}
\end{figure}

Table \ref{stab1} presents the estimated parameters for each of the three distributions. The mean directions are evidently close to one another. Figure \ref{cden} displays the spherical contour plots of the fitted densities, revealing that SESPC has more accurately captured the data's shape. The rotational symmetry assumption is rejected (p-value=0.0026), resulting in similar densities for the SIPC and SESPC, as anticipated. The densities differ, with their shapes being alike but the locations of the modes being slightly different, illustrating the effect of the elliptical symmetry present in the data. The contour lines of the ESAG distribution though are wider.

\begin{figure}[!ht]
\centering
\includegraphics[scale = 0.5, trim = 0 0 0 0]{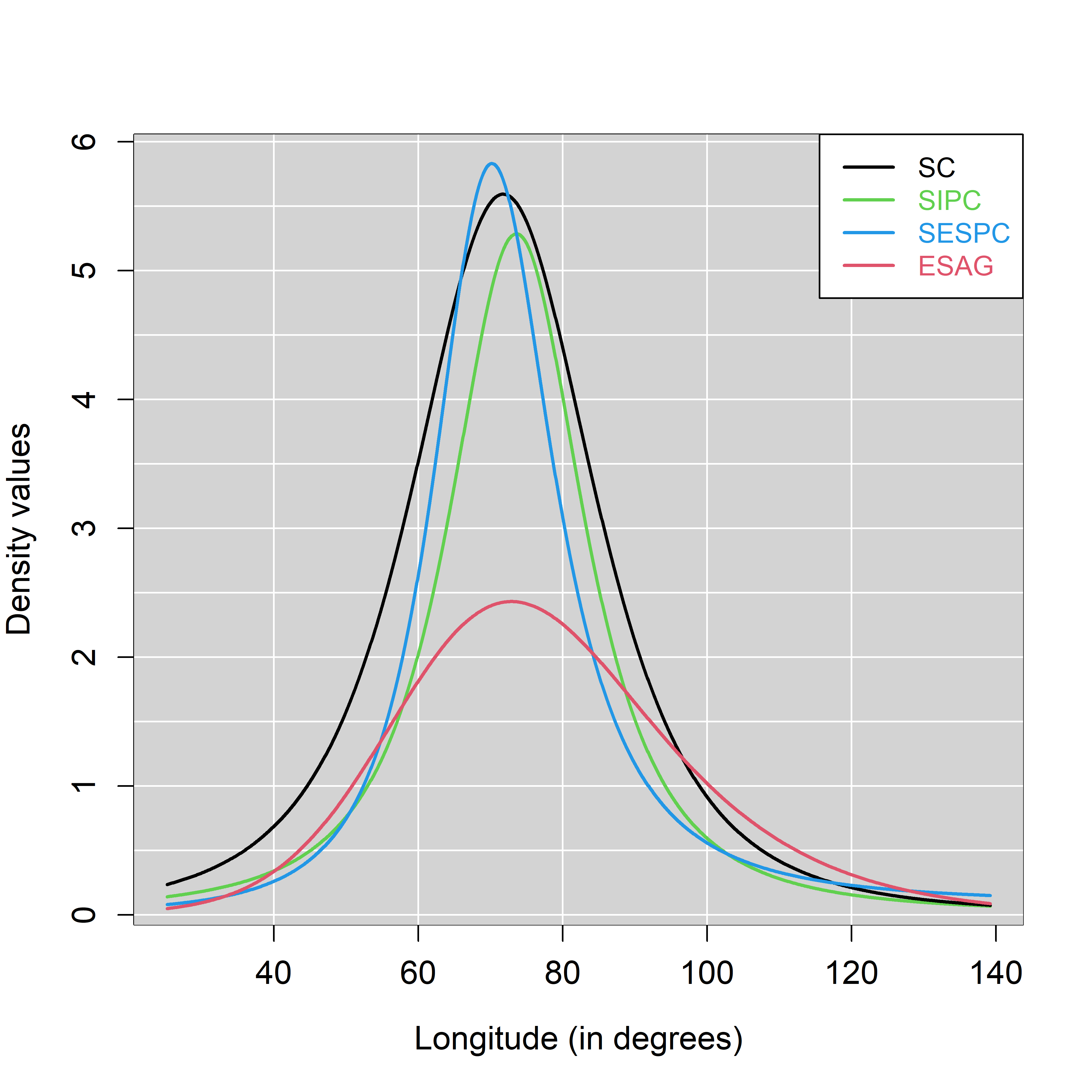}  
\caption{Comparison of the three distributions fitted to the \textit{Paleomagnetic pole} dataset: transects of the densities when Latitude=$134.71^{\circ}$.}
\label{cden2}
\end{figure}

Figure \ref{cden2} showcases the transects of the densities of the four distributions. We computed the density values for 1,000 latitude and longitude values within the observed data range. We matched the latitude to the mean direction latitude of the SESPC distribution and allowed the longitude to vary. Consequently, Figure \ref{cden2} displays a slice of the multivariate densities as a function of longitude when the latitude is $134.71^{\circ}$. 

\begin{table}[ht]
\caption{Estimated parameters for the SC, SIPC, SESPC, and ESAG distributions fitted to the \textit{Paleomagnetic pole} dataset. The columns present the estimated mean direction in Euclidean and polar coordinates, the estimated concentration parameters, and the estimated $\pmb{\theta}$ parameters for the SESPC and estimated $\pmb{\gamma}$ parameters for the ESAG distribution.}
\label{stab1}
\centering
\begin{tabular}{l|llrl} \hline
Model            & \multicolumn{3}{c}{Estimated parameters}   \\ \hline \hline
SC               & $\hat{\bm{\mu}}=(-0.699,0.224,0.679)^\top$ & $\hat{\bm{\mu}}=(134.36^{\circ}, 71.75^{\circ})^\top$ & $\hat{\rho}=0.752$ &  \\
SIPC             & $\hat{\bm{\mu}}=(-4.155,1.130,3.834)^\top$ & $\hat{\bm{\mu}}=(136.11^{\circ}, 73.58^{\circ})^\top$ & $\|\hat{\bm{\mu}}\|=5.766$ &  \\
SESPC            & $\hat{\bm{\mu}}=(-4.207,1.433,3.991)^\top$ & $\hat{\bm{\mu}}=(134.77^{\circ},70.25^{\circ})^\top$ & $\|\hat{\bm{\mu}}\|=5.973$ & $\hat{\pmb{\theta}}=(0.219,-0.846)^\top$ \\ 
ESAG             & $\hat{\bm{\mu}}=(-2.480, 1.124, 2.732)^\top$ & $\hat{\bm{\mu}}=(130.02^{\circ}, 67.63^{\circ})^\top$ & $\|\hat{\bm{\mu}}\|=3.857$ & $\hat{\pmb{\gamma}}=(0.088, -0.905)^\top$ \\
\hline \hline
\end{tabular}
\end{table}

\subsection{Spherical Data with Covariates}
The data used in this analysis are sourced from \textit{vectorcardiogram} measurements of children's heart electrical activity, taking into account different ages and genders \citep{downs1971}. Vectorcardiograms involve attaching three leads to the torso, producing a time-dependent vector that traces approximately closed curves, with each curve representing a heartbeat cycle in $\mathbb{R}^3$. A unit vector, defined as the directional component of the vector at a particular extremum across the cycles, is sometimes employed as a clinical diagnostic summary.

The dataset consists of unit vectors derived from two different lead placement systems: the McFee system (${\bf y}_i \in \mathbb{S}^2$) and the Frank system (${ \bf u}_i \in \mathbb{S}^2$) for each of 98 children from different age groups ($A_i$) and genders ($G_i$) \footnote{Both age and gender are represented by binary variables.}, with $i=1,\ldots,98$. We regressed the McFee system (${\bf y}$) on the Frank system (${\bf u}$), age ($A$), and gender ($G$) using the SIPC, SESPC, IAG and ESAG models. Table \ref{stab2} presents the estimated regression parameters for each of the four models.

The log-likelihood values for the SIPC and SESPC regression models are 40.134 and 233.467, respectively. Evidently, the isotropy assumption of the SIPC regression model is rejected. This finding is consistent with the results of \cite{paine2020}, who also rejected the isotropy assumption of the Gaussian distribution.

\begin{table}[ht]
\caption{Estimated regression parameters (their standard errors appear within parentheses) for the SIPC, SESPC, IAG and ESAG regression models fitted to the \textit{vectorcardiogram} dataset.}
\label{stab2}
\centering
\begin{tabular}{l|rrr|rrr}
\hline
          &  \multicolumn{3}{c}{SISPC} & \multicolumn{3}{c}{SESPC} \\ \hline \hline
Variables &  $Y_1$        & $Y_2$         & $Y_3$         &  $Y_1$        & $Y_2$         & $Y_3$         \\ \hline  
Constant  & -2.265(1.407) & -1.457(1.445) & -1.974(2.798) & -0.525(1.327) & 2.483(2.747)  & 1.002(1.722)  \\ 
Age       & 0.667(0.585)  & 0.791(0.685)  & 0.274(0.650)  & 0.417(0.468)  & -0.648(0.814) & -0.451(0.468) \\ 
Sex       & 1.608(0.701)  & 0.860(0.703)  & 0.483(0.731)  & 0.215(0.483)  & -1.072(0.943) & -1.365(0.508) \\ 
MQRS1     & 8.495(1.454)  & 0.282(1.060)  & 2.660(2.126)  & -2.320(0.933) & -7.100(1.882) & -4.674(1.551) \\ 
MQRS2     & -2.535(1.217) & 9.183(1.857)  & 5.540(2.196)  & 10.503(1.725) & -4.417(2.130) & 1.527(1.484)  \\ 
MQRS3     & 0.457(0.573)  & 1.024(0.640)  & 7.553(1.191)  & 1.566(0.451)  & -6.696(1.089) & 3.987(0.798)  \\ \hline
          &  \multicolumn{3}{c}{IAG} & \multicolumn{3}{c}{ESAG} \\ \hline \hline
Variables &  $Y_1$        & $Y_2$         & $Y_3$         &  $Y_1$        & $Y_2$         & $Y_3$         \\ \hline  
Constant  & 0.216(0.805)  & -0.928(0.888) & -1.784(0.972) & -0.710(1.042) & 0.001(0.717)  & -2.441(0.571) \\ 
Age       & -0.445(0.313) & -0.068(0.350) & 0.023(0.309)  & -0.692(0.442) & 0.567(0.233)  & 0.448(0.168)  \\ 
Sex       & 0.619(0.308)  & 0.837(0.353)  & 0.194(0.319)  & 0.803(0.427)  & -0.100(0.231) & -0.622(0.200) \\ 
MQRS1     & 3.983(0.626)  & 0.831(0.603)  & 2.423(0.729)  & 3.357(0.701)  & -4.338(0.578) & 0.428(0.419)  \\ 
MQRS2     & -0.812(0.698) & 4.347(0.771)  & 3.049(0.757)  & 5.361(0.939)  & 1.198(0.600)  & 3.119(0.439)  \\ 
MQRS3     & 0.637(0.400)  & 0.583(0.417)  & 3.520(0.451)  & 2.281(0.384)  & -2.066(0.362) & 2.169(0.320)  \\ \hline
\end{tabular}
\end{table}

\section{Conclusions} \label{conc}
In this study, we introduced the projected Cauchy distributions on both the circle and the sphere. For the circular case, projecting the bivariate Cauchy distribution resulted in a WC distribution with a more convenient parameterisation. The convenience pertains to the ability to have parameters that are not constrained, as is the case for the WC distribution. The concentration parameter is now located on the positive line of $\mathbb{R}$, instead within $[0, 1)$, which we think is more natural, and the location vector is unconstrained to be located in $\mathbb{R}^2$. Further, for the case of regression, in a similar manner to the SPML regression of \cite{presnell1998}, the concentration parameter does not remain constant for all observations. We then devised a generalised projected Cauchy distribution by imposing a restriction on the scatter matrix, thus adding an extra parameter. We emphasise that exploring alternative scatter matrix structures could also be valuable. Through simulation studies and real data analysis, we demonstrated that the GCPC distribution offers a superior alternative to the WC distribution, irrespective of presence of a covariate.

In the spherical case, we used a similar strategy by projecting the trivariate Cauchy distribution onto the sphere and imposing two conditions on the scatter matrix, achieving elliptical symmetry. Despite the limitations of an inconvenient density formula and challenges in generalising the density formula to higher dimensions, the projected Cauchy family of distributions provides key advantages. These benefits include a closed-form normalising constant and an efficient method of simulating values, as well as being one of the few elliptically symmetric distributions on the sphere. 

We should mention though the following point highlighted by one of the reviewers. The distributions of ${\bf Y}=\frac{\bf X}{\|{\bf X}\|}$ and ${\bf Y}^*=\frac{c{\bf X}}{\|c{\bf X}\|}=\frac{{\bf X}}{\|{\bf X}\|}$ are identical. Clearly the overparametrization is due to the fact that the distribution of $\bf Y$ is equivariant with respect to scale transformations of $\bf X$. Of course overparametrization is usually addressed by imposing constraints on the parameters, and indeed, by imposing the constraint $\pmb{\Sigma} = {\bf I}$ in the case of the CIPC and SIPC distributions and $\pmb{\Sigma}\pmb{\mu} = \pmb{\mu}$ in the case of the GCPC and SESPC distributions. Either of these constraints resolves the overparametrization by having the side-effect of fixing the dispersion of the distribution of $\bf X$: the former by completely specifying $\pmb{\Sigma}$ completely as $\bf I$, and the latter by forcing $\pmb{\Sigma}$ to have 1 as an eigenvalue.

The proposed distributions have not yet been fully explored, offering several opportunities for future research. For example, comparing the GCPC distribution with other variants derived from the projected Cauchy distribution and other three-parameter circular distributions would be valuable. Additionally, pursuing a Bayesian estimation of the PC distribution, following the approaches in \cite{wang2013} and \cite{hernandez2017}, could also provide new insights. A further direction is to discriminate between the ESAG and SESPC distributions in a formal manner, via a testing procedure, perhaps using the KLD as in \cite{dajles2022}.

\section*{Appendix}
\setcounter{section}{0}
\renewcommand{\thesubsection}{A\arabic{subsection}}
\setcounter{equation}{0}
\renewcommand{\theequation}{\thesubsection.\arabic{equation}}
\subsection{Derivation of (\ref{pc})}
We recall the terms used, $A={\bf y}^\top\pmb{\Sigma}^{-1}\pmb{\mu}$, $B={\bf y}^\top\pmb{\Sigma}^{-1}{\bf y}$ and $\Gamma^2=\pmb{\mu}^\top\pmb{\Sigma}^{-1}\pmb{\mu}$. Thus, the indefinite integral above Eq. (\ref{pc}), which we will solve in the first place, can be written as
\begin{eqnarray}
I &=& \int\frac{r}{\left(1+Br^2-2Ar+\Gamma^2\right)^{3/2}}dr \\
&=& \int\left(\frac{Br-A}{B\left(Br^2-2Ar+\Gamma^2+1\right)^{3/2}} + \frac{A}{B\left(Br^2-2Ar+\Gamma^2+1\right)^{3/2}}\right)dr \nonumber\\
&=& \frac{1}{B}\int\frac{Br-A}{\left(Br^2-2Ar+\Gamma^2+1\right)^{3/2}}dr + \frac{A}{B}\int\frac{1}{\left(Br^2-2Ar+\Gamma^2+1\right)^{3/2}}dr \nonumber \\
&=& \frac{1}{B}I_1 + \frac{A}{B}I_2 \nonumber
\end{eqnarray}
Let us now solve the first integral ($I_1$). Substitute $u=Br^2-2Ar + \Gamma^2 + 1$ and thus $du=2(Br-A)dr$, hence the first integral becomes
\begin{eqnarray}
I_1=\frac{1}{2}\int\frac{1}{u^{3/2}}du= -\frac{1}{\sqrt{u}}=-\frac{1}{\sqrt{Br^2-2Ar + \Gamma^2 + 1}}. 
\end{eqnarray}
Let us now solve the second integral ($I_2$).
\begin{eqnarray}
I_2=B^{3/2}\int\frac{1}{\left[\left(Br-A\right)^2+B\left(\Gamma^2+1\right)-A^2\right]^{3/2}}dr=B^{3/2}I_3.
\end{eqnarray}
Substitute $u=Br-A$ and hence $du=Bdr$, thus $I_3$ can be written as
\begin{eqnarray}
I_3=\frac{1}{B} \int\frac{1}{\left[u^2+B\left(\Gamma^2+1\right)-A^2\right]^{3/2}}du=\frac{1}{B}I_4.
\end{eqnarray}
Again, using substitution, $u=\sqrt{B\left(\Gamma^2+1\right)-A^2}\tan(v)$ and $v=\arctan\left(\frac{u}{\sqrt{B\left(\Gamma^2+1\right)-A^2}}\right)$. Then \\
$du=\sqrt{B\left(\Gamma^2+1\right)-A^2}\sec^2(v)dv$. Thus, $I_4$ becomes
\begin{eqnarray}
I_4 &=& \int\frac{\sqrt{B\left(\Gamma^2+1\right)-A^2}\sec^2(v)}{\left[\left(B\left(\Gamma^2+1\right)-A^2\right)\tan^2(v)+B\left(\Gamma^2+1\right)-A^2\right]^{3/2}}dv \nonumber \\
&=&\frac{1}{B\Gamma^2+B-A^2}\int\frac{1}{sec(v)}dv= \frac{sin(v)}{B\Gamma^2+B-A^2}. 
\end{eqnarray}
We undo the last substitution, and hence $\sin(v)=\sin\left[\arctan\left(\frac{u}{\sqrt{B\left(\Gamma^2+1\right)-A^2}}\right)\right]=\frac{u}{\sqrt{B\left(\Gamma^2+1\right)-A^2}\sqrt{\frac{u^2}{\sqrt{B\left(\Gamma^2+1\right)-A^2}}+1}}$. We plug this last result into $I_4$ to obtain 
\begin{eqnarray}
I_4=\frac{u}{\left(B\Gamma^2+B-A^2\right)\sqrt{B\left(\Gamma^2+1\right)-A^2}\sqrt{\frac{u^2}{\sqrt{B\left(\Gamma^2+1\right)-A^2}}+1}}   
\end{eqnarray}
and hence $I_3$ becomes
\begin{eqnarray}
I_3=\frac{u}{B\left(B\Gamma^2+B-A^2\right)\sqrt{B\left(\Gamma^2+1\right)-A^2}\sqrt{\frac{u^2}{\sqrt{B\left(\Gamma^2+1\right)-A^2}}+1}}. 
\end{eqnarray}
We now undo the substitution that took us from $I_2$ to $I_3$ and obtain
\begin{eqnarray}
I_2=\frac{B^{3/2}\left(Br-A\right)}{B\left(B\Gamma^2+B-A^2\right)\sqrt{B\left(\Gamma^2+1\right)-A^2}\sqrt{\frac{\left(Br-A\right)^2}{\sqrt{B\left(\Gamma^2+1\right)-A^2}}+1}}.
\end{eqnarray}
Finally, $I$ is written as 
\begin{eqnarray}
\begin{split}
I&=-\frac{1}{B}\frac{1}{\sqrt{Br^2-2Ar + \Gamma^2 + 1}} + \\
&\frac{A}{B}\frac{B^{3/2}\left(Br-A\right)}{B\left(B\Gamma^2+B-A^2\right)\sqrt{B\left(\Gamma^2+1\right)-A^2}\sqrt{\frac{\left(Br-A\right)^2}{\sqrt{B\left(\Gamma^2+1\right)-A^2}}+1}} + c,
\end{split}
\end{eqnarray}
where $c$ is a constant. After some rearrangement, the above integral becomes
\begin{eqnarray} 
I=\frac{Ar-\Gamma^2-1}{\left(B\Gamma^2+B-A^2\right)\sqrt{Br^2-2Ar+\Gamma^2+1}} + c.    
\end{eqnarray}
Hence the definite integral is equal to
\begin{eqnarray}
I &=& \int_0^{\infty}\frac{r}{\left(1+Br^2-2Ar+\Gamma^2\right)^{3/2}}dr = \frac{B\sqrt{\Gamma^2+1}+A\sqrt{B}}{B\left(B\Gamma^2+B-A^2\right)}.
\end{eqnarray}
By simplifying the above expression using identities and by adding the ignored constant terms, we end up with the expression in (\ref{pc}). 

\subsection{Derivatives of the log-likelihood of the CIPC distribution with the parameterisation of (\ref{cipc})}
\begin{eqnarray*}
\ell &=& - \sum_{i=1}^n\log\left(\sqrt{\gamma^2+1}-\alpha_i\right) -n \log\left(2\pi\right) \\
\frac{\partial\ell}{\partial \pmb{\mu}} &=&  - \sum_{i=1}^n\dfrac{\frac{\pmb{\mu}}{\sqrt{\gamma^2+1}}-{\bf y}_i}{\sqrt{\gamma^2+1}-\alpha_i} \\
\frac{\partial^2\ell}{\partial \pmb{\mu}\pmb{\mu}^\top} &=& - \sum_{i=1}^n\dfrac{\frac{{\bf I}_2\sqrt{\gamma^2+1}-\frac{\pmb{\mu}\pmb{\mu}^\top}{\sqrt{\gamma^2+1}}}{\gamma^2+1}\left(\sqrt{\gamma^2+1}-\alpha_i\right)-\left(\frac{\pmb{\mu}}{\sqrt{\gamma^2+1}}-{\bf y}_i\right)\left(\frac{\pmb{\mu}}{\sqrt{\gamma^2+1}}-{\bf y}_i\right)^\top}{\left(\sqrt{\gamma^2+1}-\alpha_i\right)^2}.
\end{eqnarray*}

\subsection{Derivatives of the log-likelihood of the CIPC distribution with the parameterisation of  (\ref{cipc2})}
\begin{eqnarray*}
\ell &=&  \sum_{i=1}^n\log\left(\sqrt{\gamma^2+1}-\gamma\cos\left(\theta_i-\omega\right)\right) -n \log\left(2\pi\right) \\
\frac{\partial\ell}{\partial \omega} &=&  \sum_{i=1}^n\dfrac{\gamma\sin\left(\theta_i-\omega\right)}{\sqrt{\gamma^2+1} - \gamma\cos\left(\theta_i-\omega\right)} 
\\
\frac{\partial\ell}{\partial \gamma} &=&  - \sum_{i=1}^n\dfrac{\frac{\gamma}{\sqrt{\gamma^2+1}}-\cos\left(\theta_i-\omega\right)}{\sqrt{\gamma^2+1}-\gamma\cos\left(\theta_i-\omega\right)} \\
\frac{\partial^2\ell}{\partial \omega^2} &=&
\sum_{i=1}^n\dfrac{\gamma^2\sin^2\left(\theta_i-\omega\right)}{\left(\sqrt{\gamma^2+1}-\gamma\cos\left(\theta_i-\omega\right)\right)^2}-\dfrac{\gamma\cos\left(\theta_i-\omega\right)}{\sqrt{\gamma^2+1} - \gamma\cos\left(\theta_i-\omega\right)}  \\
\frac{\partial^2\ell}{\partial \gamma^2} &=&
\sum_{i=1}^n \dfrac{\left(\frac{\gamma}{\sqrt{\gamma^2+1}}-\cos\left(\theta_i-\omega\right)\right)^2}{\left(\sqrt{\gamma^2+1}-\gamma\cos\left(\theta_i-\omega\right)\right)^2}-\dfrac{\frac{1}{\sqrt{\gamma^2+1}}-\frac{\gamma^2}{\left(\gamma^2+1\right)^\frac{3}{2}}}{\sqrt{\gamma^2+1}-\gamma\cos\left(\theta_i-\omega\right)} \\
\frac{\partial^2\ell}{\partial \omega \partial \gamma}  &=& \sum_{i=1}^n\dfrac{\sin\left(\theta_i-\omega\right)}{\sqrt{\gamma^2+1}\left(\sqrt{\gamma^2+1}-\gamma\cos\left(\theta_i-\omega\right)\right)^2}.
\end{eqnarray*}

\subsection{Derivatives of the log-likelihood of the GCPC distribution using the parametrisation in (\ref{gcpc3})}
\begin{tiny}
\begin{eqnarray*}
\ell &=& -\frac{n}{2}\log(\rho)-\sum_{i=1}^n\log\left(\cos^2(\theta_i-\omega)\right) - \frac{1}{2}\sum_{i=1}^n\log\left(1+\frac{1+\tan^2(\theta_i-\omega)}{\rho}\right) \\ 
& & -\sum_{i=1}^n\log\left[\sqrt{\left(\gamma^2+1\right)\left(1+\frac{1+\tan^2(\theta_i-\omega)}{\rho}\right)}-\gamma\right] \\
\frac{\partial\ell}{\partial \omega} &=& -\sum_{i=1}^n\left[\frac{\sqrt{\gamma^2+1}\sec^2\left(\theta_i-\omega\right)\tan\left(\theta_i-\omega\right)}{\rho\sqrt{\frac{\tan^2\left(\theta_i-\omega\right)+\rho}\rho}\left(\sqrt{\gamma^2+1}\sqrt{\frac{\tan^2\left(\theta_i-\omega\right)+\rho}{\rho}}-\gamma\right)} -2\tan\left(\theta_i-\omega\right)+\frac{\sec^2\left(\theta_i-\omega\right)\tan\left(\theta_i-\omega\right)}{\tan^2\left(\theta_i-\omega\right)+\rho} \right] \\
\frac{\partial^2\ell}{\partial \omega^2} &=& 
-\sum_{i=1}^n\left[ \frac{2\sqrt{\gamma^2+1}\sec^2\left(\theta_i-\omega\right)\tan^2\left(\theta_i-\omega\right)}{\rho\sqrt{\frac{\tan^2\left(\theta_i-\omega\right)+1}{\rho}+1}\left(\sqrt{\gamma^2+1}\sqrt{\frac{\tan^2\left(\omega-\theta_i\right)+1}{\rho}+1}-\gamma\right)} + \frac{2\sec^2\left(\theta_i-\omega\right)\tan^2\left(\omega-\theta_i\right)}{{\rho}\left(\frac{\tan^2\left(\theta_i-\omega\right)+1}{\rho}+1\right)} \right]\\
& & -\sum_{i=1}^n \left[ \frac{\sqrt{\gamma^2+1}\sec^4\left(\theta_i-\omega\right)}{\rho\sqrt{\frac{\tan^2\left(\theta_i-\omega\right)+1}\rho+1}\left(\sqrt{\gamma^2+1}\sqrt{\frac{\tan^2\left(\theta_i-\omega\right)+1}{\rho}+1}-\gamma\right)} + \frac{\sec^4\left(\theta_i-\omega\right)}{\rho\left(\frac{\tan^2\left(\theta_i-\omega\right)+1}{\rho}+1\right)} \right] \\
& & +\sum_{i=1}^n \left[ \frac{\sqrt{\gamma^2+1}\sec^4\left(\theta_i-\omega\right)\tan^2\left(\theta_i-\omega\right)}{\rho^2\left(\frac{\tan^2\left(\theta_i-\omega\right)+1}{\rho}+1\right)^\frac{3}{2}\left(\sqrt{\gamma^2+1}\sqrt{\frac{\tan^2\left(\omega-\theta_i\right)+1}\rho+1}-\gamma\right)}+\frac{2\sec^4\left(\theta_i-\omega\right)\tan^2\left(\theta_i-\omega\right)}{{\rho}^2\left(\frac{\tan^2\left(\theta_i-\omega\right)+1}{\rho}+1\right)^2} \right] \\
& & +\sum_{i=1}^n \left[ \frac{\left(\gamma^2+1\right)\sec^4\left(\theta_i-\omega\right)\tan^2\left(\theta_i-\omega\right)}{\rho^2\left(\frac{\tan^2\left(\theta_i-\omega\right)+1}{\rho}+1\right)\left(\sqrt{\gamma^2+1}\sqrt{\frac{\tan^2\left(\theta_i-\omega\right)+1}{\rho}+1}-\gamma\right)^2} +2\tan^2\left(\theta_i-\omega\right)+2 \right]  \\
\frac{\partial\ell}{\partial \gamma} &=& -\sum_{i=1}^n\frac{\frac{\sqrt{\frac{\tan^2\left(\theta-\omega\right)+1}{\rho}+1}\,\gamma}{\sqrt{\gamma^2+1}}-1}{\sqrt{\frac{\tan^2\left(\theta_i-\omega\right)+1}{\rho}+1}\sqrt{\gamma^2+1}-\gamma} \\
\frac{\partial^2\ell}{\partial \gamma^2} &=& \sum_{i=1}^n \left[ \frac{\left(\frac{\sqrt{\frac{\tan^2\left(\theta_i-\omega\right)+1}{\rho}+1}\,\gamma}{\sqrt{\gamma^2+1}}-1\right)^2}{\left(\sqrt{\frac{\tan^2\left(\theta_i-\omega\right)+1}{\rho}+1}\sqrt{\gamma^2+1}-\gamma\right)^2}-\frac{\frac{\sqrt{\frac{\tan^2\left(\theta_i-\omega\right)+1}{\rho}+1}}{\sqrt{\gamma^2+1}}-\frac{\sqrt{\frac{\tan^2\left(\theta_i-\omega\right)+1}{\rho}+1}\,\gamma^2}{\left(\gamma^2+1\right)^\frac{3}{2}}}{\sqrt{\frac{\tan^2\left(\theta_i-\omega\right)+1}{\rho}+1}\sqrt{\gamma^2+1}-\gamma} \right]
\end{eqnarray*}
\end{tiny}
\begin{tiny}
\begin{eqnarray*}
\frac{\partial\ell}{\partial \rho} &=& -\frac{1}{2}\sum_{i=1}^n \left[\frac{\sqrt{\gamma^2+1}\left(\frac{1}{\rho}-\frac{\rho+\tan^2\left(\theta_i-\omega\right)+1}{\rho^2}\right)}{\sqrt{\frac{\rho+\tan^2\left(\theta_i-\omega\right)+1}{\rho}}\left(\sqrt{\gamma^2+1}\sqrt{\frac{\rho+\tan^2\left(\theta_i-\omega\right)+1}{\rho}}-\gamma\right)}+\frac{\rho\left(\frac{1}{\rho}-\frac{\rho+\tan^2\left(\theta_i-\omega\right)+1}{\rho^2}\right)}{\rho+\tan^2\left(\theta_i-\omega\right)+1}\right] - \frac{n}{2\rho} \\
\frac{\partial^2\ell}{\partial \rho^2} &=& \sum_{i=1}^n \left[ \frac{\sqrt{\gamma^2+1}\left(\frac{1}{\rho}-\frac{\rho+\tan^2\left(\theta_i-\omega\right)+1}{\rho^2}\right)^2}{4\left(\frac{\rho+\tan^2\left(\theta_i-\omega\right)+1}{\rho}\right)^\frac{3}{2}\,\left(\sqrt{\gamma^2+1}\sqrt{\frac{\rho+\tan^2\left(\theta_i-\omega\right)+1}{\rho}}-\gamma\right)} -\frac{\frac{1}{\rho}-\frac{\rho+\tan^2\left(\theta_i-\omega\right)+1}{\rho^2}}{2\left(\rho+\tan^2\left(\theta_i-\omega\right)+1\right)} \right] \\
& & -\sum_{i=1}^n \left[ \frac{\sqrt{\gamma^2+1}\left(\frac{2\left(\rho+\tan^2\left(\theta_i-\omega\right)+1\right)}{\rho^3}-\frac{2}{\rho^2}\right)}{2 \sqrt{\frac{\rho+\tan^2\left(\theta_i-\omega\right)+1}{\rho}}\left(\sqrt{\gamma^2+1}\sqrt{\frac{\rho+\tan^2\left(\theta_i-\omega\right)+1}{\rho}}-\gamma\right)} - \frac{\rho\left(\frac{1}{\rho}-\frac{\rho+\tan^2\left(\theta_i-\omega\right)+1}{\rho^2}\right)}{2\left(\rho+\tan^2\left(\theta_i-\omega\right)+1\right)^2} \right] \\
& & +\sum_{i=1}^n \frac{\left(\gamma^2+1\right){\rho}\left(\frac{1}{\rho}-\frac{\rho+\tan^2\left(\theta_i-\omega\right)+1}{\rho^2}\right)^2}{4\left(\rho+\tan^2\left(\theta_i-\omega\right)+1\right)\left(\sqrt{\gamma^2+1}\sqrt{\frac{\rho+\tan^2\left(\theta_i-\omega\right)+1}{\rho}}-\gamma\right)^2} \\
& & -\sum_{i=1}^n \frac{\rho\left(\frac{2\left(\rho+\tan^2\left(\theta_i-\omega\right)+1\right)}{\rho^3}+\frac{2}{\rho^2}\right)}{2\left(\rho+\tan^2\left(\theta_i-\omega\right)+1\right)}+\frac{n}{2\rho^2} \\
\frac{\partial^2\ell}{\partial \omega \partial \gamma} &=& 
\frac{\sec^2\left(\theta_i-\omega\right)\tan\left(\theta_i-\omega\right)}{\rho\sqrt{\frac{\tan^2\left(\theta_i-\omega\right)+1}{\rho}+1}\sqrt{\gamma^2+1}\left(\sqrt{\frac{\tan^2\left(\theta_i-\omega\right)+1}{\rho}+1}\sqrt{\gamma^2+1}-\gamma\right)^2} \\
\frac{\partial^2\ell}{\partial \omega \partial \rho} &=& 
\sum_{i=1}^n \frac{\sec^2\left(\theta_i-\omega\right)\tan\left(\theta_i-\omega\right)\sqrt{\gamma^2+1}\left(-\frac{\sec^2\left(\theta_i-\omega\right)\,\gamma}{2\sqrt{\frac{\sec^2\left(\theta_i-\omega\right)}{\rho}+1}\,\rho}+\gamma\sqrt{\frac{\sec^2\left(\theta_i-\omega\right)}{\rho}+1}-\sqrt{\gamma^2+1}\right)}{\left(\gamma\sqrt{\frac{\sec^2\left(\theta_i-\omega\right)}{{\rho}}+1}\,\rho-\sqrt{\gamma^2+1}\,\rho-\sec^2\left(\theta_i-\omega\right)\sqrt{\gamma^2+1}\right)^2} \\
& & -\sum_{i=1}^n \frac{\sec^2\left(\theta_i-\omega\right)\tan\left(\theta_i-\omega\right)}{\left(\rho+\sec^2\left(\theta_i-\omega\right)\right)^2} \\
\frac{\partial^2\ell}{\partial \gamma \partial \rho} &=& 
\sum_{i=1}^n \frac{\tan^2\left(\theta_i-\omega\right)+1}{2\rho^2\sqrt{\gamma^2+1}\left(\sqrt{\gamma^2+1}\sqrt{\frac{\tan^2\left(\theta_i-\omega\right)+1}{\rho}+1}-\gamma\right)^2\sqrt{\frac{\tan^2\left(\theta_i-\omega\right)+1}{\rho}+1}}.
\end{eqnarray*}
\end{tiny}
\subsection{A note on the cumulative probability function of the GCPC distribution}
By using (\ref{gcpc3}) we may compute its cumulative probability function to be
\begin{eqnarray*}
P(\theta_L \leq \theta \leq \theta_U) &=& \frac{1}{\pi}\arctan\left[\frac{\sqrt{\rho}\left(\sqrt{\gamma^2+1}+\gamma\right)\left(\sqrt{1 + \frac{\tan^2\left(\theta_U-\omega\right)}{\rho}}-1\right)}{\tan\left(\theta_U-\omega\right)}\right] \\ 
& & - \frac{1}{\pi}\arctan\left[\frac{\sqrt{\rho}\left(\sqrt{\gamma^2+1}+\gamma\right)\left(\sqrt{1 + \frac{\tan^2\left(\theta_L-\omega\right)}{\rho}}-1\right)}{\tan\left(\theta_L-\omega\right)}\right].
\end{eqnarray*}
The problem with this formula is that it is valid only when $\pi/2 < \theta-\omega < \pi/2$.


\begin{thebibliography}{}

\bibitem[\protect\citeauthoryear{Abe and Pewsey}{Abe and Pewsey}{2011}]{abe2011}
Abe, T. and A.~Pewsey (2011).
\newblock Sine-skewed circular distributions.
\newblock {\em Statistical Papers\/}~{\em 52\/}(3), 683--707.

\bibitem[\protect\citeauthoryear{Agostinelli and Lund}{Agostinelli and Lund}{2017}]{circular2017}
Agostinelli, C. and U.~Lund (2017).
\newblock {\em {R package \texttt{circular}: Circular Statistics (version 0.4-93)}}.
\newblock {https://CRAN.R-project.org/package=circular}.

\bibitem[\protect\citeauthoryear{Bullock, Feix and Dollar}{Bullock, Feix and Dollar}{2014}]{bullock2014}
Bullock, I.~M., T. Feix and A.~M. Dollar (2014).
\newblock {Analyzing human fingertip usage in dexterous precision manipulation: Implications for robotic finger design}.
\newblock {\em 2014 IEEE/RSJ International Conference on Intelligent Robots and Systems\/}, 1622--1628.

\bibitem[\protect\citeauthoryear{Chang}{Chang}{1986}]{chang1986}
Chang~T. (1986).
\newblock {Spherical regression}.
\newblock {\em The Annals of Statistics\/}~{\em 14\/}(3), 907--924.

\bibitem[\protect\citeauthoryear{Dajles and Cavanaugh}{Dajles and Cavanaugh}{2022}]{dajles2022}
Dajles~A. and Cavanaugh~J. (2022).
\newblock {Probabilistic Pairwise Model Comparisons Based on Bootstrap479
Estimators of the Kullback–Leibler Discrepancy}.
\newblock {\em Entropy\/}~{\em 24\/}, 1483.

\bibitem[\protect\citeauthoryear{Dietrich and Richter}{Dietrich and
Richter}{2017}]{dietrich2017}
Dietrich, T. and W.~D.~Richter (2017).
\newblock {Classes of geometrically generalized von Mises distributions}.
\newblock {\em Sankhya B\/}~{\em 79\/}(1), 21--59.

\bibitem[\protect\citeauthoryear{Downs, Liebam and Mackay}{Downs, Liebam and Mackay}{1971}]{downs1971}
Downs, T. and J.~Liebman and W.~Mackay (1971).
\newblock {Statistical methods for vectorcardiogram orientations}.
\newblock {\em Vectorcardiography\/}~{\em 2\/} 216--222.

\bibitem[\protect\citeauthoryear{Fisher}{Fisher}{1995}]{fisher1995}
Fisher, N.~I. (1995).
\newblock {\em Statistical analysis of circular data}.
\newblock Cambridge University Press.

\bibitem[\protect\citeauthoryear{Fisher}{Fisher}{1953}]{fisher1953}
Fisher, R.~A. (1953).
\newblock {Dispersion on a sphere}.
\newblock {\em Proceedings of the Royal Society of London. Series A.
Mathematical and Physical Sciences\/}~{\em 217\/}(1130), 295--305.

\bibitem[\protect\citeauthoryear{Gatto and Jammalamadaka}{Gatto and
Jammalamadaka}{2007}]{gatto2007}
Gatto, R. and S.~R. Jammalamadaka (2007).
\newblock {The generalized von Mises distribution}.
\newblock {\em Statistical Methodology\/}~{\em 4\/}(3), 341--353.

\bibitem[\protect\citeauthoryear{Gill and Hangartner}{Gill and Hangartner}{2010}]{gill2010}
Gill, J. and D.~Hangartner (2010).
\newblock {Circular data in political science and how to handle it}.
\newblock {\em Political Analysis\/}~{\em 18\/}(3), 316--336.

\bibitem[\protect\citeauthoryear{Heaton et al.}{Heaton et al.}{2014}]{heaton2014}
Heaton, M.~J., M. Katzfuss, C. Berrett and D.~W. Nychka (2014).
\newblock {Constructing valid spatial processes on the sphere using kernel convolutions}.
\newblock {\em Environmetrics\/}~{\em 25\/}(1), 2--15.

\bibitem[\protect\citeauthoryear{Hernandez-Stumpfhauser, Breidt and van der Woerd}{Hernandez-Stumpfhauser, Breidt and van der Woerd}{2017}]{hernandez2017}
Hernandez-Stumpfhauser, D., F.~J. Breidt and M.~J. van der Woerd (2017).
\newblock {The general projected normal distribution of arbitrary dimension: Modeling and Bayesian inference}.
\newblock {\em Bayesian Analysis\/}~{\em 12\/}(1), 113--133.

\bibitem[\protect\citeauthoryear{Horne et al.}{Horne et al.}{2007}]{horne2007}
Horne, J.~S., E.~O. Garton, S.~M. Krone and J.~S. Lewis (2010).
\newblock {Analyzing animal movements using Brownian bridges}.
\newblock {\em Ecology\/}~{\em 88\/}(9), 2354--2363.

\bibitem[\protect\citeauthoryear{Jammalamadaka}{Jammalamadaka and Sengupta}{2001}]{jammalamadaka2001}
Jammalamadaka, R.~S. and A.~Sengupta. (2001).
\newblock {\em Topics in circular statistics}.
\newblock World Scientific, Singapore.

\bibitem[\protect\citeauthoryear{Jones and Pewsey}{Jones and
Pewsey}{2005}]{jones2005}
Jones, M. and A.~Pewsey (2005).
\newblock {A family of symmetric distributions on the circle}.
\newblock {\em Journal of the American Statistical Association\/}~{\em
100\/}(472), 1422--1428.

\bibitem[\protect\citeauthoryear{Jones and Pewsey}{Jones and
Pewsey}{2012}]{jones2012}
Jones, M. and A.~Pewsey (2012).
\newblock {Inverse Batschelet distributions for circular data}.
\newblock {\em Biometrics\/}~{\em 68\/}(1), 183--193.

\bibitem[\protect\citeauthoryear{Kato and Jones}{Kato and
Jones}{2010}]{kato2010}
Kato, S. and M.~Jones (2010).
\newblock {A family of distributions on the circle with links to, and
applications arising from, M{\"o}bius transformation}.
\newblock {\em Journal of the American Statistical Association\/}~{\em
105\/}(489), 249--262.

\bibitem[\protect\citeauthoryear{Kato and Jones}{Kato and
Jones}{2013}]{kato2013}
Kato, S. and M.~Jones (2013).
\newblock {An extended family of circular distributions related to wrapped
Cauchy distributions via Brownian motion}.
\newblock {\em Bernoulli\/}~{\em 19\/}(1), 154--171.

\bibitem[\protect\citeauthoryear{Kato and McCullagh}{Kato and
McCullagh}{2020}]{kato2020}
Kato, S. and P.~McCullagh (2020).
\newblock {Some properties of a Cauchy family on the sphere derived from the
M{\"o}bius transformations}.
\newblock {\em Bernoulli\/}~{\em 26\/}(4), 3224--3248.

\bibitem[\protect\citeauthoryear{Kendall}{Kendall}{1974}]{kendall1974}
Kendall, D.~G. (1974).
\newblock {Pole-seeking Brownian motion and bird navigation}.
\newblock {\em Journal of the Royal Statistical Society: Series B
(Methodological)\/}~{\em 36\/}(3), 365--402.

\bibitem[\protect\citeauthoryear{Kent}{Kent}{1982}]{kent1982}
Kent, J.~T. (1982).
\newblock {The Fisher-Bingham distribution on the sphere}.
\newblock {\em Journal of the Royal Statistical Society: Series B
(Methodological)\/}~{\em 44\/}(1), 71--80.

\bibitem[\protect\citeauthoryear{Kent, Hussein and Jah}{Kent, Hussein and Jah}{2016}]{kent2016}
Kent, J.~T., I. Hussein and M.~K. Jah (2016).
\newblock {Directional distributions in tracking of space debris}.
\newblock {\em 2016 19th International Conference on Information Fusion (FUSION)\/}, 2081--2086.

\bibitem[\protect\citeauthoryear{Kim and SenGupta}{Kim and
SenGupta}{2013}]{kim2013}
Kim, S. and A.~SenGupta (2013).
\newblock {A three-parameter generalized von Mises distribution}.
\newblock {\em Statistical Papers\/}~{\em 54\/}(3), 685--693.

\bibitem[\protect\citeauthoryear{Landler, Ruxton and Malkemper}{Landler, Ruxton and Malkemper}{2018}]{landler2018}
Landler, L. and D.~G. Ruxton and E.~P. Malkemper (2018).
\newblock {Circular data in biology: advice for effectively implementing statistical procedures}.
\newblock {\em Behavioral Ecology and Sociobiology\/}~{\em 72\/}, 1--10.

\bibitem[\protect\citeauthoryear{Mardia and Jupp}{Mardia and
Jupp}{2000}]{mardia2000}
Mardia, K. and P.~Jupp (2000).
\newblock {\em Directional {S}tatistics}.
\newblock John {W}iley \& {S}ons.

\bibitem[\protect\citeauthoryear{Mardia}{Mardia}{1975}]{mardia1975}
Mardia, K.~V. (1975).
\newblock Statistics of directional data.
\newblock {\em Journal of the Royal Statistical Society: Series B
(Methodological)\/}~{\em 37\/}(3), 349--371.

\bibitem[\protect\citeauthoryear{Mardia}{Mardia}{1972}]{mardia1972}
Mardia, K.~V. (1972).
\newblock {\em Statistics of directional data}.
\newblock Academic Press, London, UK.

\bibitem[\protect\citeauthoryear{Nelder and Mead}{Nelder and Mead}{1965}]{nelder1965}
Nelder, J.~A. and R.~Mead (1975).
\newblock A simplex method for function minimization.
\newblock {\em The Computer Journal\/}~{\em 7\/}(4), 308--313.

\bibitem[\protect\citeauthoryear{Nolan}{Nolan}{2021}]{nolan2021}
Nolan~P. (2021).
\newblock {R package \texttt{SphericalCubature}: Numerical Integration over Spheres and
Balls in n-Dimensions; Multivariate Polar Coordinates (version 1.5)}.
\newblock {https://CRAN.R-project.org/package=SphericalCubature}.

\bibitem[\protect\citeauthoryear{Nu{\~n}ez-Antonio and
Guti{\'e}rrez-Pe{\~n}a}{Nu{\~n}ez-Antonio and Guti{\'e}rrez-Pe{\~n}a}{2005}]{nunez2005}
Nu{\~n}ez-Antonio, G. and E.~Guti{\'e}rrez-Pe{\~n}a (2005).
\newblock {A Bayesian analysis of directional data using the projected normal
distribution}.
\newblock {\em Journal of Applied Statistics\/}~{\em 32\/}(10), 995--1001.

\bibitem[\protect\citeauthoryear{Oliveira, Crujeiras and Rodr{\'i}guez-Casal}{2014}]{npcirc2014}
Oliveira, M., R.~M.~Crujeiras and A.~Rodr{\'i}guez-Casal (2014).
\newblock {NPCirc: An R Package for Nonparametric Circular Methods}.
\newblock {\em Journal of Statistical Software\/}~{\em 61\/}(9), 1--26.

\bibitem[\protect\citeauthoryear{Oliveira, Crujeiras and Rodr{\'\i}guez-Casal}{2014}]{oliveira2014}
Oliveira, M., R.~M.~Crujeiras and A.~Rodr{\'\i}guez-Casal (2014).
\newblock {CircSiZer: an exploratory tool for circular data}.
\newblock {\em Environmental and Ecological Statistics\/}~{\em 21\/}(10), 143--159.

\bibitem[\protect\citeauthoryear{Paine, Preston, Tsagris and Wood}{Paine
et~al.}{2018}]{paine2018}
Paine, P., S.~P. Preston, M.~Tsagris and A.~T. Wood (2018).
\newblock {An elliptically symmetric angular Gaussian distribution}.
\newblock {\em Statistics and Computing\/}~{\em 28\/}(3), 689--697.

\bibitem[\protect\citeauthoryear{Paine, Preston, Tsagris and Wood}{Paine
et~al.}{2020}]{paine2020}
Paine, P., S.~P. Preston, M.~Tsagris and A.~T. Wood (2020).
\newblock {Spherical regression models with general covariates and anisotropic errors}.
\newblock {\em Statistics and Computing\/}~{\em 30\/}(1), 153--165.

\bibitem[\protect\citeauthoryear{Pewsey}{Pewsey}{2000}]{pewsey2000}
Pewsey, A. (2000).
\newblock The wrapped skew-normal distribution on the circle.
\newblock {\em Communications in Statistics--Theory and Methods\/}~{\em
29\/}(11), 2459--2472.

\bibitem[\protect\citeauthoryear{Pewsey}{Pewsey}{2008}]{pewsey2008}
Pewsey, A. (2008).
\newblock The wrapped stable family of distributions as a flexible model for circular data.
\newblock {\em Computational Statistics \& Data Analysis\/}~{\em 52\/}(3),
1516--1523.

\bibitem[\protect\citeauthoryear{Pewsey, Lewis and Jones}{Pewsey
et~al.}{2007}]{pewsey2007}
Pewsey, A., T.~Lewis and M.~Jones (2007).
\newblock The wrapped t family of circular distributions.
\newblock {\em Australian \& New Zealand Journal of Statistics\/}~{\em
49\/}(1), 79--91.

\bibitem[\protect\citeauthoryear{Presnell, Morrison and Littell}{Presnell
et~al.}{1998}]{presnell1998}
Presnell, B., S.~P. Morrison and R.~C. Littell (1998).
\newblock Projected multivariate linear models for directional data.
\newblock {\em Journal of the American Statistical Association\/}~{\em
93\/}(443), 1068--1077.

\bibitem[\protect\citeauthoryear{Schmidt}{Schmidt}{1976}]{schmidt1976}
Schmidt, P. (1976).
\newblock {The non-uniqueness of the Australian Mesozoic palaeomagnetic pole position}.
\newblock {\em Geophysical Journal International\/}~{\em 47\/}(2), 285--300.

\bibitem[\protect\citeauthoryear{Shirota and Gelfand}{Shirota and Gelfand}{2017}]{shirota2017}
Shirota, S. and A.~E. Gelfand (2017).
\newblock {Space and circular time log Gaussian Cox processes with application to crime event data}.
\newblock {\em The Annals of Applied Statistics\/}~{\em 11\/}(2), 481--503.

\bibitem[\protect\citeauthoryear{Soler et~al.}{Soler et~al.}{2019}]{soler2019}
Soler, J.~D and H. Beuther, M. Rugel, Y. Wang, P.~C. Clark, S.~.C.~O Glover, P.~F. Goldsmith, M. Heyer, L.~D. Anderson, A. Goodman and others. (2019).
\newblock {Histogram of oriented gradients: a technique for the study of molecular cloud formation}.
\newblock {\em Astronomy \& Astrophysics\/}~{\em 622}, A166.

\bibitem[\protect\citeauthoryear{Straub et~al.}{Straub et~al.}{2015}]{straub2015}
Straub J., J. Chang, O. Freifeld and J. Fisher III (2015).
\newblock {A Dirichlet process mixture model for spherical data}
\newblock {\em Artificial Intelligence and Statistics\/}, 930--938.

\bibitem[\protect\citeauthoryear{Tsagris et~al.}{Tsagris et~al.}{2023}]{directional2023}
Tsagris M., G.~Athineou, C.~Adam, A.~Sajib, E.~Amson and M.~J. Waldstein (2023).
\newblock {R package \texttt{Directional}: A Collection of Functions for Directional Data Analysis (version 6.4)}.
\newblock {https://CRAN.R-project.org/package=Directional}.

\bibitem[\protect\citeauthoryear{von Mises}{von Mises}{1918}]{von1918}
von Mises, R. (1918).
\newblock {Uber die" Ganzzahligkeit" der Atomgewicht und verwandte Fragen}.
\newblock {\em Physikal. Z.\/}~{\em 19}, 490--500.

\bibitem[\protect\citeauthoryear{Wang and Gelfand}{Wang and
Gelfand}{2013}]{wang2013}
Wang, F. and A.~E. Gelfand (2013).
\newblock Directional data analysis under the general projected normal distribution.
\newblock {\em Statistical Methodology\/}~{\em 10\/}(1), 113--127.

\end{thebibliography}
\end{document}